\documentclass[12pt,a4paper]{article}

\usepackage{graphics}{}
\pdfoutput=1
\usepackage[pdftex]{}
\usepackage{amsmath,amssymb,amsfonts}

\usepackage[dvips]{epsfig}
\usepackage{multicol}
\usepackage{xspace}
\usepackage{url}
\usepackage{hyperref}
\usepackage{amsmath}

\newcommand{\eq}[1]{\begin{align} #1 \end{align}}

\begin{document}

\vspace{1.8cm}
\begin{center}
{\Large \bf Onset of deconfinement in nucleus-nucleus collisions:}
\end{center}
\begin{center}
{\Large \bf Review for pedestrians and experts}
\end{center}

\vspace*{0.8cm}
\begin{center}
Marek Gazdzicki$^{a,b}$, Mark Gorenstein$^{c,d}$ and Peter Seyboth$^{e,b}$
\end{center}

\vspace*{0.2cm}
\begin{center}
{\it $^{a}$Institut f\"ur Kernphysik, University of Frankfurt, Frankfurt, Germany\\}
{\it $^{b}$Jan Kochanowski University, Kielce, Poland\\}
{\it $^{c}$Bogolyubov Institute for Theoretical Physics, Kiev, Ukraine\\}
{\it $^{d}$Frankfurt Institute for Advanced Studies, Frankfurt, Germany\\}
{\it $^{e}$Max-Planck-Institut fuer Physik, Munich, Germany\\}
\end{center}

\vspace*{0.8cm}
\begin{abstract}
Evidence for the energy threshold of creating the quark-gluon
plasma in nucleus-nucleus collisions, the so-called onset of
deconfinement, has been found by the energy scan program of the
NA49 experiment at the CERN SPS. In this paper we review the
experimental and theoretical status of this phenomenon. First, the
basic, qualitative ideas are presented for non-experts. Next, the
latest experimental results are compared to a statistical model
within which the onset of deconfinement and its signals had been
predicted. Finally, alternative interpretations and open questions
are discussed.

\vspace*{2.0cm}

\end{abstract}

\thispagestyle{empty}

\newpage

\tableofcontents

\newpage

\newpage
\section{Introduction}

One of the important issues of contemporary physics is the
understanding of strong interactions and in particular the study
of the properties of strongly interacting matter in equilibrium.
What are the phases of this matter and what do the transitions
between them look like~? These questions motivate broad
experimental and theoretical efforts since more than 40 years. The
study of high energy collisions between two atomic nuclei give us
the unique possibility to address these issues in well controlled
laboratory experiments. In particular, the advent of the quark
model of hadrons and the development of the commonly accepted
theory of strong interactions, quantum chromodynamics (QCD),
naturally led to expectations that matter at very high densities
may exist in a state of quasi-free quarks and gluons, the quark-gluon
plasma (QGP)~\cite{qgp0,qgp1,qgp2}.

Experimental searches for QGP signals started at the Super Proton
Synchrotron (SPS) of the European Organization for Nuclear
Research (CERN) and the Alternating Gradient Synchrotron (AGS) of
Brookhaven National Laboratory (BNL) in the mid 1980s. Today they
are pursued also at much higher collision energies at the
Relativistic Heavy Ion Collider (RHIC) at BNL. Soon experiments on
nucleus-nucleus collisions at the Large Hadron Collider (LHC) in
CERN will join the world effort at energies 20 times higher than
at RHIC. Most probably that the QGP is formed at the early stage
of heavy ion collisions at the top SPS energy and at RHIC
energies. Unambiguous evidence of the QGP state was however
missing. This may be attributed to the difficulty of obtaining
unique and quantitative predictions of the expected QGP signals
from the theory of strong interactions.

For this reason the NA49 Collaboration at the CERN SPS has
searched over the past years for signs of the onset of QGP
creation in the energy dependence of hadron production properties.
This search was motivated by a statistical model~\cite{GaGo}
showing that the onset of deconfinement should lead to rapid
changes of the energy dependence of numerous experimentally
detectable properties of the collisions, all appearing in a common
energy domain. The predicted features have recently been
observed~\cite{evidence} and dedicated experiments now continue
detailed studies in the energy region of the onset of
deconfinement.

It is thus time for a summary. In this paper we review the
experimental and theoretical status of the onset of deconfinement.
First, the basic qualitative ideas are presented for
non-experts. Next, a quantitative model within which the onset
of deconfinement and its signals were predicted is reexamined
and compared with the latest experimental results. Finally,
alternative interpretations and open questions are discussed.

\section{Onset of Deconfinement for Pedestrians}

Phase transitions are fascinating physical phenomena.
Small changes in temperature or pressure
lead to dramatic changes in macroscopic properties of
matter.
Common examples from our daily life are transitions
between solids, liquids and gases like boiling and
freezing of water.
\begin{figure}[!htb]
\begin{center}
\begin{minipage}[b]{1.0\linewidth}
\includegraphics[width=1.0\linewidth]{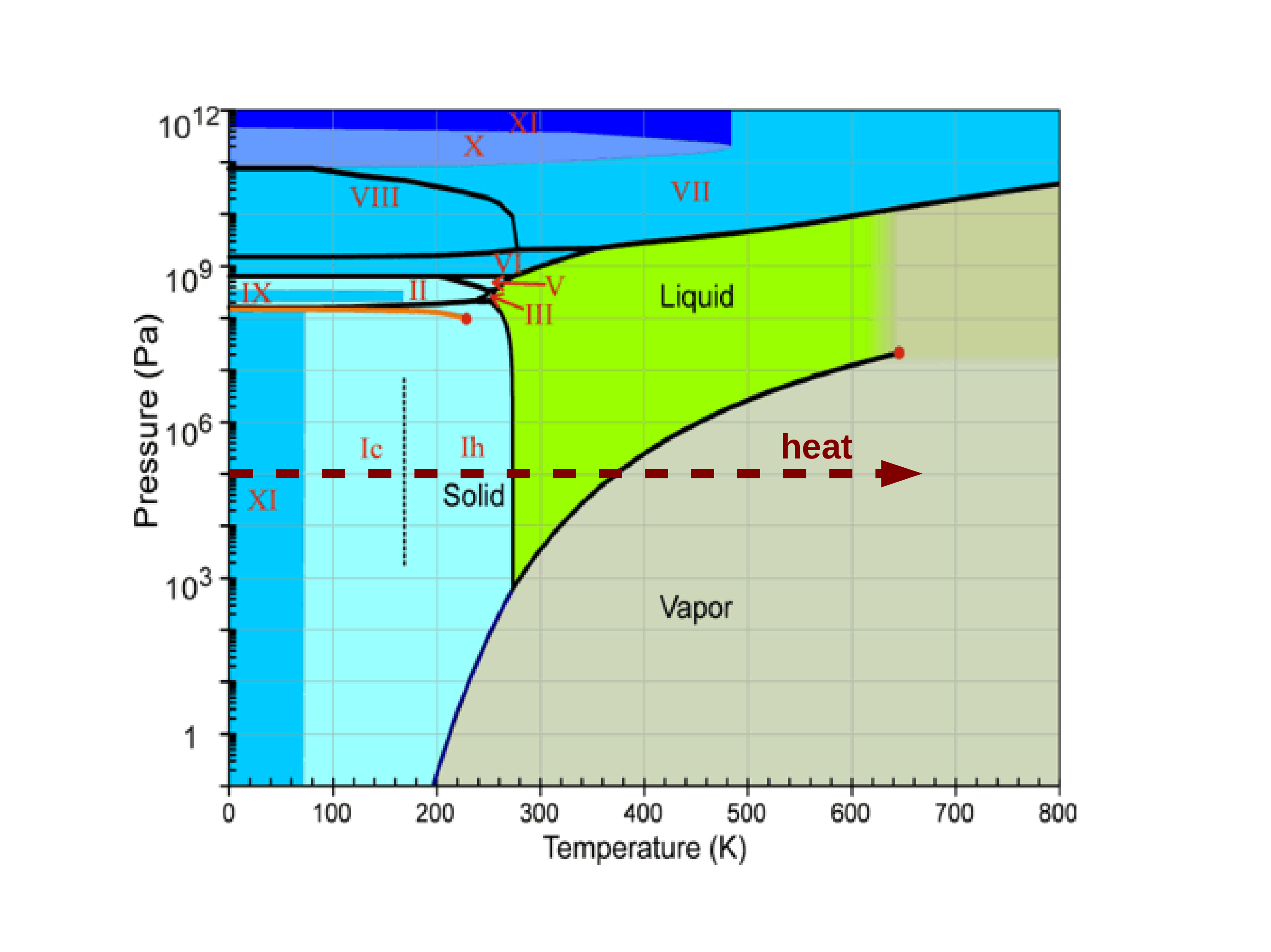}
\end{minipage}

\vspace{-0.5cm}
\caption{\label{water_phases}
Phases of water. When adding heat (energy) at constant pressure water
is transformed from solid to liquid and then from liquid to vapor as
indicated by the dashed arrow.
}
\end{center}
\end{figure}
The well known phase diagram of water is shown in
Fig.~\ref{water_phases}, where the regions of existence of the
various phases of water are depicted in a diagram of pressure and
temperature. When adding heat to water one increases its
temperature moving through its different phases and crossing their
boundaries, as indicated by the dashed arrow in
Fig.~\ref{water_phases} for the example of constant atmospheric
pressure. Dependence of the water temperature on the amount of
added heat, called the heating curve of water, is shown in
Fig.~\ref{water_heating_curve}. In pure phases, such as ice, water
or vapor, the temperature increases monotonically with added heat.
The two regions of constant temperature (steps) signal the
ice-water and water-vapor phase transitions. In these mixed phase
regions added heat is used for the phase transformation instead of
the increase of temperature as in the pure phase regions.
\begin{figure}[!htb]
\begin{center}
\begin{minipage}[b]{1.0\linewidth}
\includegraphics[width=1.0\linewidth]{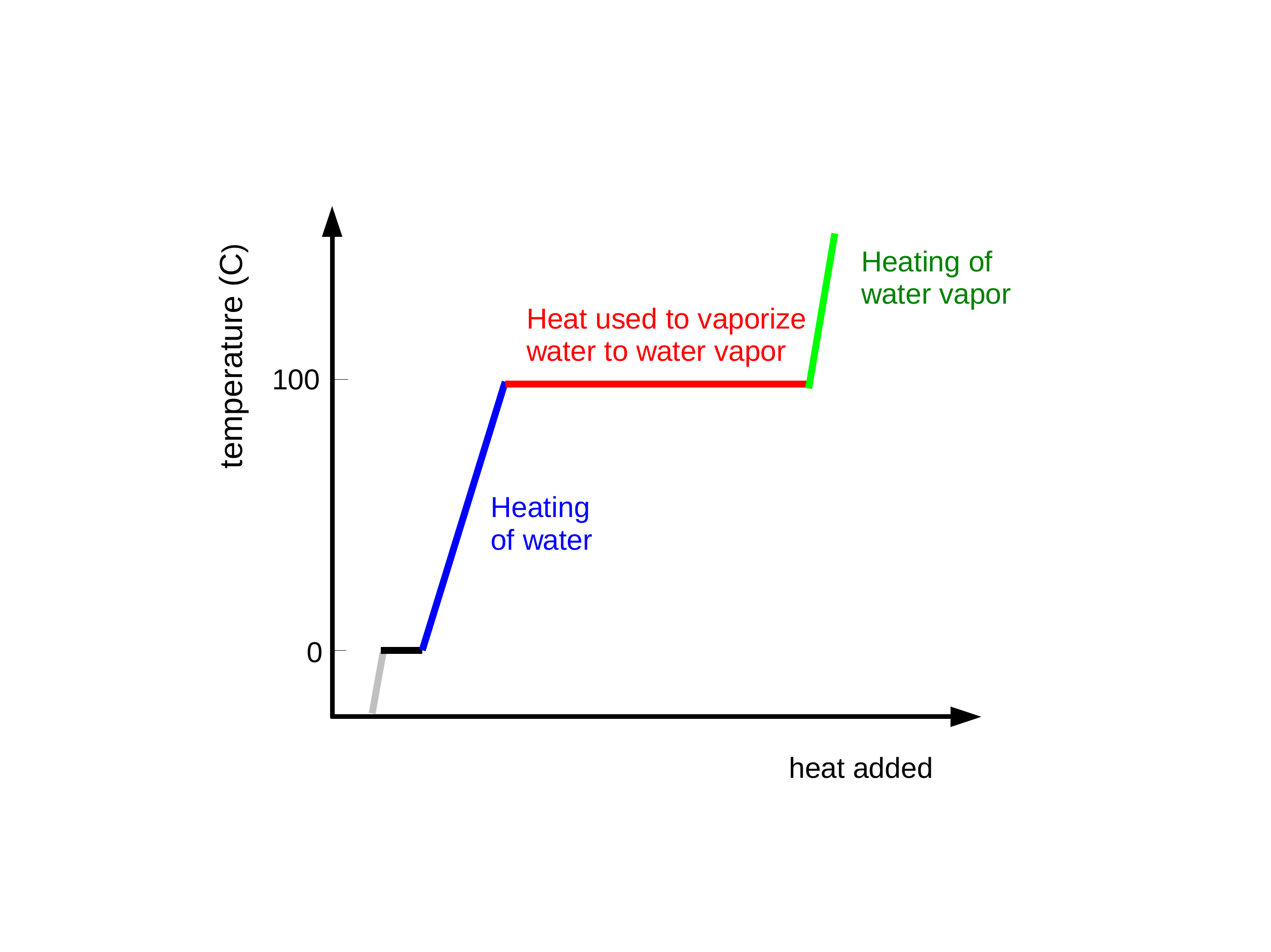}
\end{minipage}

\vspace{-2cm}
\caption{\label{water_heating_curve}
Heating curve of water at fixed atmospheric pressure. It
corresponds to the trajectory in the phase diagram of
water indicated by the dashed arrow in Fig.~~\ref{water_phases}.
}
\end{center}
\end{figure}

Properties of water and other substances surrounding us and the
transitions between their various phases are determined by
electromagnetic interactions of atoms and molecules. On the other
hand, the properties of atomic nuclei
which are built from nucleons (protons and neutrons) are determined
by strong interactions.
Naturally, the question arises whether strongly interacting matter
also exists in distinct phases. What are their properties? At
which temperatures  do the transitions between them take place?
What do these transitions look like?

Since more than 40 years it is known that hadrons
(i.e. mesons and baryons; all strongly interacting particles
observed in nature are called hadrons)
consist of more elementary particles, the quarks and gluons.
However, isolated quarks or gluons were never observed.
They seem to be always confined in the interior of hadrons.
But could a different phase of strongly interacting matter
exist in which quarks and gluons are deconfined~?

\begin{figure}[!htb]
\begin{center}
\begin{minipage}[b]{1.0\linewidth}
\includegraphics[width=1.0\linewidth]{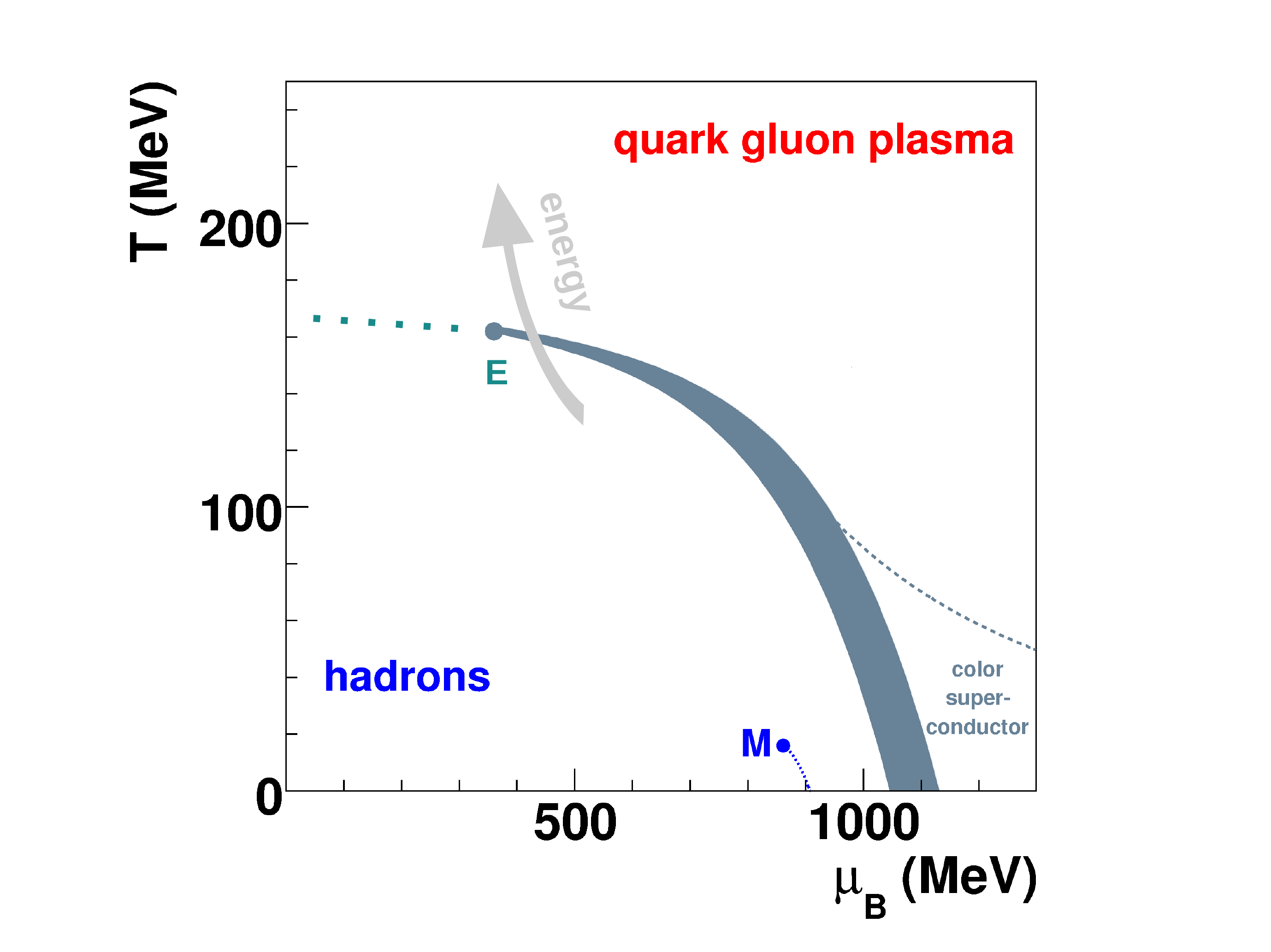}
\end{minipage}
\caption{\label{sim_phases}
Phases of strongly interacting matter.
With increasing collision energy the matter created at the early
stage of nucleus-nucleus collisions changes its properties as
indicated by the arrow. At low energy it is in the confined phase
(hadrons), at sufficiently high energy in the deconfined phase
(QGP). $M$ is the critical point of the nuclear liquid-gas phase
transition. The shaded band shows the 1$^{st}$ order phase
boundary between the hadron and QGP phase which is expected to end
in a critical end point $E$. At $E$ the sharp phase transition
turns into a rapid crossover indicated as the dotted line. }
\end{center}
\end{figure}

There are 3 parameters
which describe the thermodynamical properties of a system. In
non-relativistic systems they are temperature, particle number
density, and pressure. The equation of state connects them, e.g.,
the pressure is a well defined function of temperature and
particle density for a specific substance. In experiments on water
one can most easily fix temperature and pressure to define the point on
the phase diagram in Fig.~\ref{water_phases}. Unlike in water, the
number of particles is not conserved in strongly interacting
relativistic matter.  Instead of particle number density 
the baryonic number, i.e. the difference between the number of
baryons and anti-baryons, is conserved. In calculations it is
convenient to use the equivalent variables baryonic number density 
or baryonic chemical potential.
The phase diagram of strongly interacting matter emerging from
theoretical considerations and experimental results is shown in
Fig.~\ref{sim_phases} in terms of the commonly used variables temperature
and baryonic chemical potential. Laboratory experiments
(see discussion below) can create strongly interacting matter with
different temperatures $T$ and baryonic chemical potentials
$\mu_B$. The functional dependence of the pressure on  $T$ and $\mu_B$, i.e.
the equation of state of strongly interacting matter, remains the
subject of intensive experimental and theoretical studies.

A transition between the deconfined and the confined phase of
strongly interacting matter probably took place during the
expansion and cooling of the early Universe, about 1~microsecond
after the Big Bang. Cosmological signatures of this transition are
difficult to identify today. However, extremely dense strongly
interacting matter fills the interior of neutron stars.
%
%
Arguments in favor of the existence of quark matter in the center
of such stars were advanced already in the 1960s, soon after
formulation of the quark hypothesis. One of the pioneering
papers~\cite{qgp1} argued: ``A neutron has a radius  of about
0.5--1~fm (1~fm = 10$^{-15}$~m), and so has a density of about
8$\cdot$10$^{14}$ g$\cdot$cm$^{-3}$, whereas the central density of a
neutron star can be as much as $10^{16}-10^{17}$ g$\cdot$cm$^{-3}$. In
this case, one must expect the hadrons to overlap, and their
individuality to be confused. Therefore, we suggest that matter at
such densities is a quark soup.'' The creation of matter in a
deconfined phase, i.e. in the QGP phase, may be the only
possibility to `see' quarks and gluons moving freely in a large
volume.

\begin{figure}[!htb]
\begin{center}
\begin{minipage}[b]{1.0\linewidth}
\includegraphics[width=1.0\linewidth]{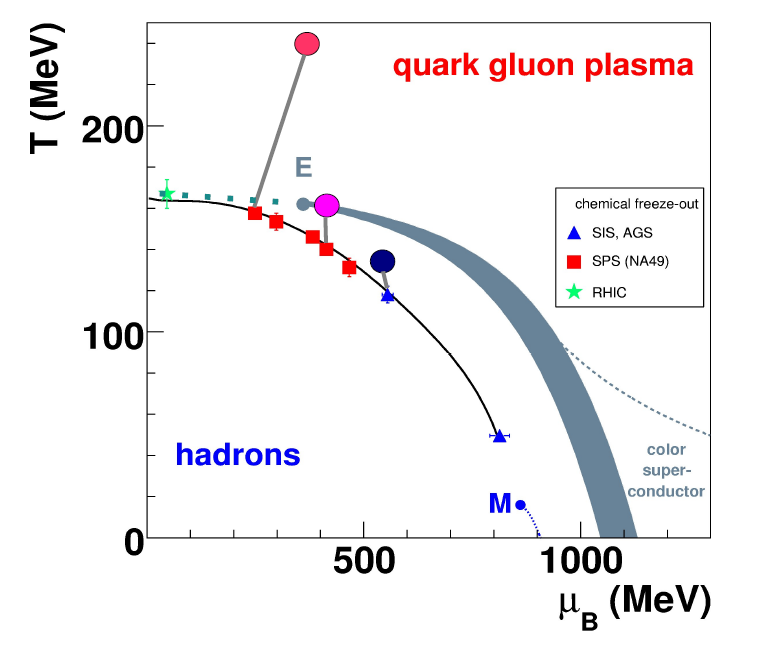}
\end{minipage}
\caption{\label{sim_phases1} Parameters of strongly interacting
matter created at the early stage of nucleus-nucleus interactions
are shown by the full circles for central Pb+Pb (Au+Au) collisions
at the top AGS energy ($\sqrt{s}_{NN} \approx 5.5$~GeV),
intermediate SPS energy ($\sqrt{s}_{NN} \approx 7.6$~GeV) and top
SPS energy ($\sqrt{s}_{NN} \approx 17$~GeV). The created fireball expands
and cools along trajectories indicated by solid lines and
decouples at the freeze-out points (full squares, triangles and
star). }
\end{center}
\end{figure}

Cosmological and  astrophysical objects with the required properties
are, unfortunately, difficult to investigate.
Systematic study of the properties of strongly interacting matter
requires a method to create it under well controlled conditions in
the laboratory. The study of collisions of two  heavy nuclei gives
us this possibility. Such a collision produces a droplet of
strongly interacting matter of high energy density, the so-called
fireball.
It is natural to expect that with increasing collision energy the
fireball energy density also increases. Thus like in the case of
water heating and observing  successive transitions between its
phases, we hope that with increasing collision energy we can
detect anomalies in the energy dependence of hadron production
properties and thus discover successive transitions between
various phases of strongly interacting matter created at the early
stage of  collisions. The arrow in Fig.~\ref{sim_phases}
schematically traces the position of the initially created
fireball on the phase diagram when the energy of nucleus-nucleus
collisions is increasing.
At sufficiently high collision energy this matter droplet may
reach the QGP phase (see Fig.~\ref{sim_phases}). Unfortunately the
life time of the fireball is very short, about $10^{-22}$~seconds.
It quickly expands, cools down (see Fig.~\ref{sim_phases1}) and
finally decays into hadrons and a few light nuclei. These decay
products are measured in detectors surrounding the collision
point.

The first phase transition of strongly interacting matter was
observed studyingg  collisions at very low energies~\cite{liquide}
(the energy per nucleon-nucleon pair in the center of mass system $\sqrt{s}_{NN} < 2$~GeV). 
This transition between a nuclear liquid and
a nuclear gas happens at a temperature of about 6$\cdot$10$^{10}$ K
(5~MeV). The phase transition line and critical point $M$ lie at
large $\mu_B$ and small $T$ inside the hadron phase region of the phase diagram
as shown in Figs.~\ref{sim_phases} and~\ref{sim_phases1}.

Emerging results from the study of high energy collisions of nuclei
confirm the existence of the second phase
transition in strongly interacting matter which was suggested by QCD.
It is the so--called deconfinement phase transition.
Let us briefly present how one looked for this phase transition,
and explain the main results.

\begin{figure}[!htb]
\begin{center}
\begin{minipage}[b]{1.0\linewidth}
\includegraphics[width=0.5\linewidth]{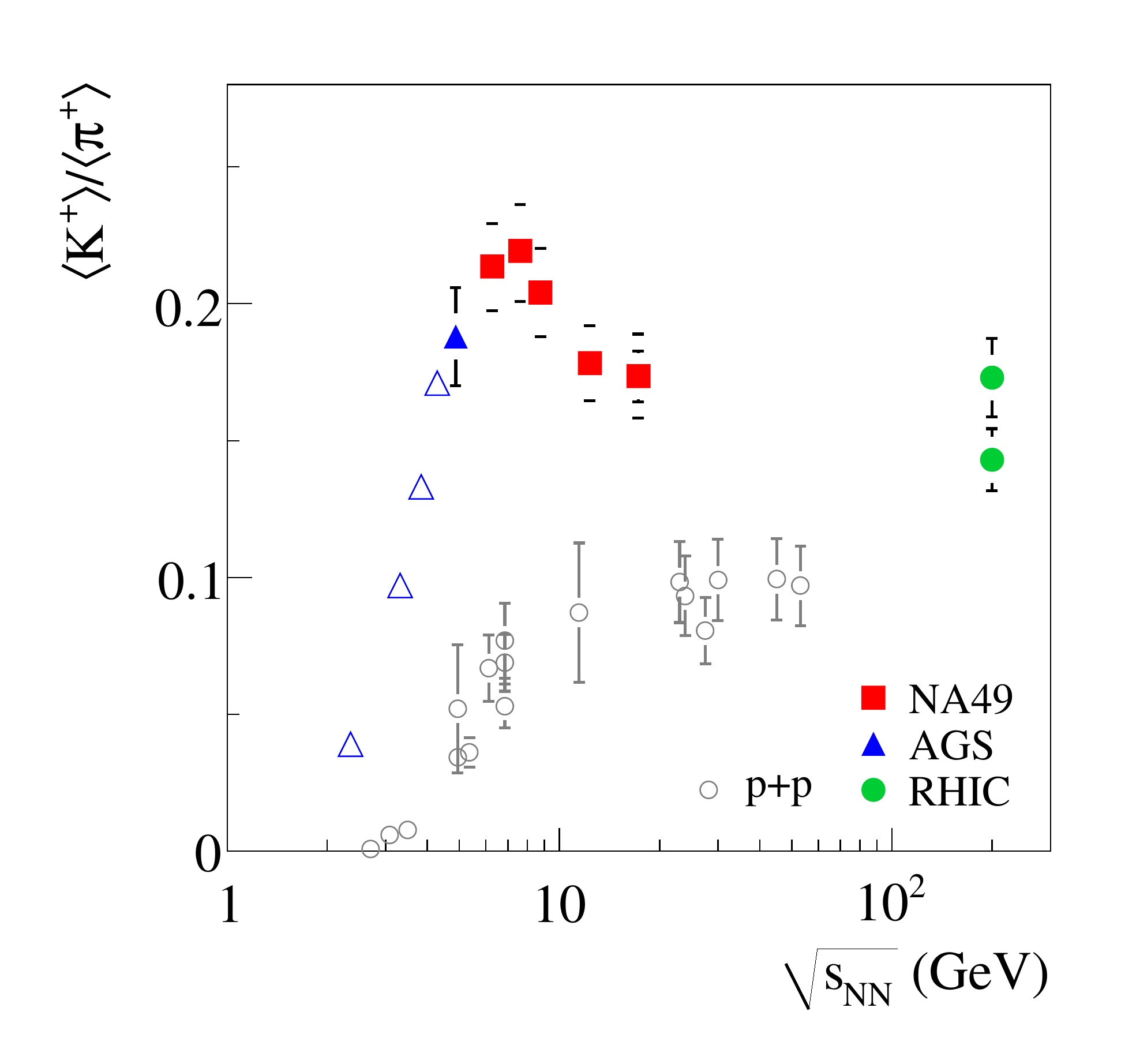}
\includegraphics[width=0.5\linewidth]{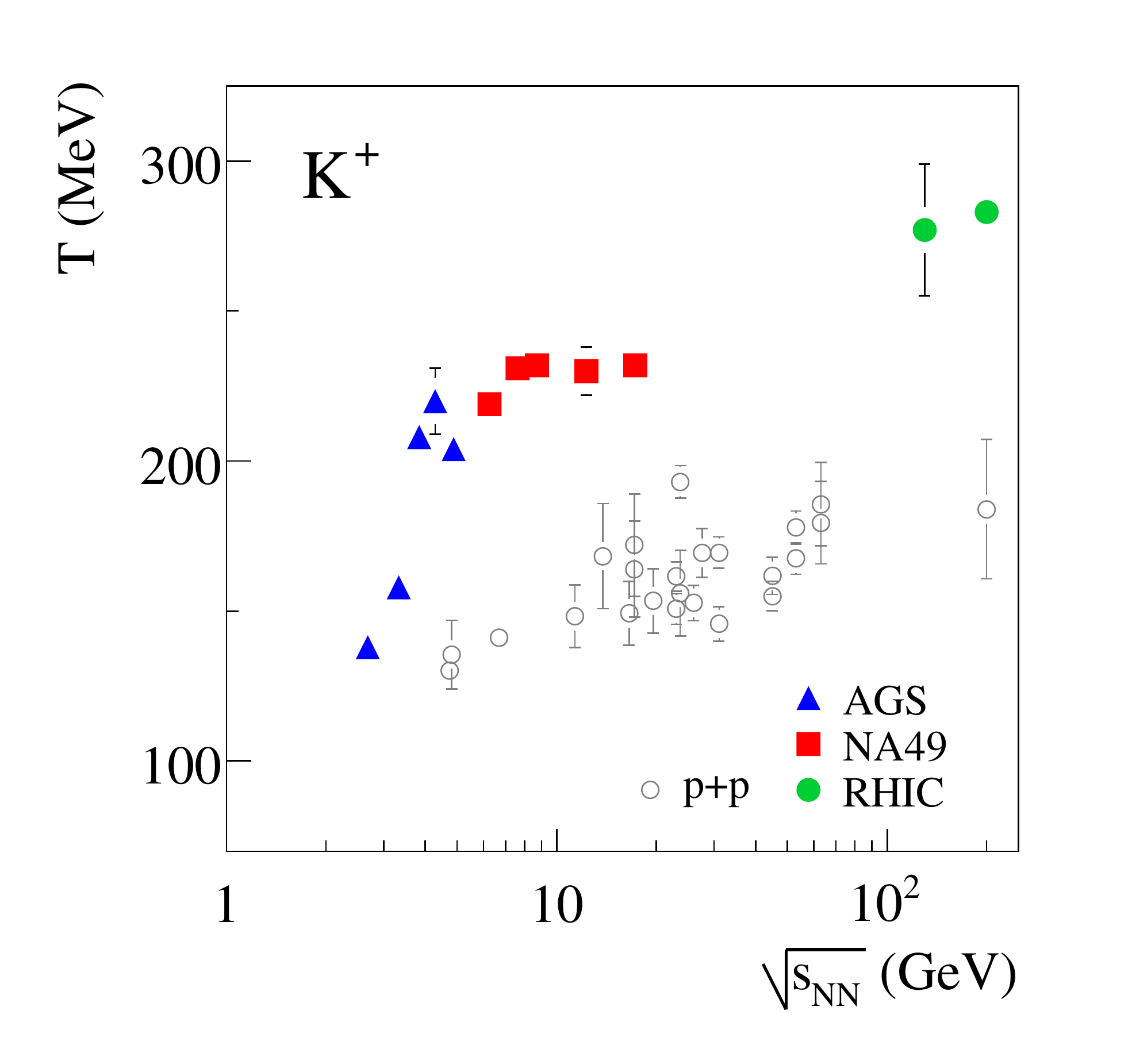}
\end{minipage}
\caption{\label{sim_heating_curve}
Heating curves of strongly interacting matter.
Hadron production properties (see text for details) are plotted
as a function of collision energy for central Pb+Pb (Au+Au)
collisions (upper set of points) and p+p interactions
(lower set of points)~\cite{evidence}
}
\end{center}
\end{figure}

The search for the phase of deconfined strongly interacting
matter, the  QGP, already has a long history. It received a boost
from the first acceleration of oxygen and sulfur nuclei at the
CERN SPS in 1986 ($\sqrt{s}_{NN} \approx 20$~GeV) and of lead
nuclei in 1994 ($\sqrt{s}_{NN} \approx 17$~GeV). Measurements from
an array of experiments indicated that the critical energy density
was probably exceeded and matter with unusual properties appeared
to be formed in the early stage of the collisions~\cite{cernpr}. A
key problem was the identification of experimental signatures of
QGP creation. Several signatures of the formation of a transient
QGP state during the early stage of the collision had been
proposed in the past~\cite{Rafelski, Satz}. However, the
uniqueness of these signatures came under renewed scrutiny and
they were found not to be specific for the creation of QGP (see
Appendix~8.1 for details).

In the mid 1990s a study of results from experiments at CERN and
the AGS at BNL (maximum energy $\sqrt{s}_{NN} \approx 5.5$ GeV)
raised~\cite{GaRo1, GaRo2} intriguing questions concerning  the
energy dependence of hadron production between top AGS and SPS
energies. In response to these questions a statistical model of
the early stage of the collision process was proposed~\cite{GaGo}
in which an equation of state with a 1$^{st}$ order phase
transition was assumed. In this model the onset of
deconfinement led to the prediction of a non-monotonic collision
energy dependence of several hadron production properties. In
particular, the model predicted a sharp maximum in the ratio of
multiplicities of strange hadrons (the hadrons which contain $s$
and $\overline{s}$ quarks) to pions (the lightest hadron) at the
beginning of the transition region, at about $\sqrt{s}_{NN}
\approx 7.5$ GeV. This prediction triggered an extension of the
experimental program at the SPS, the energy 
scan program~\cite{na49scan}. Within
this program head--on (central) collisions of two lead nuclei
(Pb+Pb) were registered at several lower SPS energies 
($\sqrt{s}_{NN} = $ 6.3, 7.6, 8.7 and 12.3~GeV) by the NA49 experiment.
Other heavy ion experiments at the SPS (NA45, NA50, NA57 and NA60)
participated in selected runs of this program~\cite{othersinscan}. 
Final results,
obtained mainly by the NA49 collaboration, confirm the qualitative
expectations and the quantitative predictions of the model: rapid
changes in properties of hadron production occur within a narrow
energy range, $\sqrt{s}_{NN}=$~7--12~GeV~\cite{evidence}.

The most dramatic effect is seen in the energy dependence of the
ratio of total particle yields of kaons and pions, $\langle K^+
\rangle/\langle \pi^+ \rangle$, in central Pb+Pb collisions which
is plotted in Fig.~\ref{sim_heating_curve}~$left$. Following a
fast threshold rise the ratio passes through a sharp maximum in
the SPS range and then seems to settle to a plateau value at
higher energies. Kaons are the lightest strange hadrons and due to
approximate isospin symmetry the $\langle K^+ \rangle$ yield
counts about half of the strange (anti-)quarks produced in the
collisions and contained in the reaction products (see
Appendix~8.2 for details). Thus
Fig.~\ref{sim_heating_curve}~$left$ demonstrates that the fraction
of strangeness carrying particles in the produced matter passes
through a sharp maximum at the SPS in nucleus--nucleus collisions.
This feature is not observed for proton--proton reactions as shown
by the open dots in Fig.~\ref{sim_heating_curve}~$left$.

A second important result is the stationary value of the apparent
temperature $T$ of $K^+$ mesons in central Pb+Pb collisions at SPS
energies as shown in  Fig.~\ref{sim_heating_curve}~$right$. In the
fireball picture the apparent temperature is related to the local
random motion of the particles and their collective expansion
velocity in the direction transverse to the collision axis.

Presently the sharp maximum and the following plateau in the
energy dependence of the $\langle K^+ \rangle/\langle \pi^+
\rangle$ ratio has only been reproduced by the statistical model
of the early stage. In this model it reflects the decrease in the
ratio of strange to non-strange number of degrees of freedom when
deconfinement sets in.
The stationary value of the apparent temperature  of $K^+$ mesons
was predicted~\cite{van-hove,Hu:95,GoGaBu} as a consequence of the
constant pressure and temperature  at
the early stage of nucleus--nucleus collisions in the SPS energy range
due to the coexistence of hadronic and deconfined phases.

{\bf These results serve as evidence that the deconfinement phase
transition in Pb+Pb collisions starts in the SPS energy range.}
%
%
The exciting and rich physics which can be studied in
nucleus-nucleus collisions at the CERN SPS energies motivates
ongoing and future experimental programs at the CERN
SPS~\cite{Gazdzicki:2006fy,proposal}, BNL RHIC~\cite{rhic_low},
FAIR SIS~\cite{cbm} and JINR NICA~\cite{mpd}.

\section{Statistical Model of the Early Stage (SMES)}

The experimental search for the onset of deconfinement performed
by the experiment NA49 at the CERN SPS was motivated by the
predictions of the Statistical Model of the Early
Stage~\cite{GaGo}, which treats the creation of the fireball in
nucleus--nucleus collisions in a statistical model approach. The
model does not attempt a description of the subsequent
(hydro--)dynamical evolution of the fireball. First, this section
reviews the main assumptions of the originally formulated model
and presents results obtained from analytical and numerical calculations
as well as comparison to the experimental data available 12 years ago, 
when the model was formulated.
Next, later extensions of the SMES are discussed which address
collective flow  at freeze-out~\cite{GoGaBu,bleicher} and
event-by-event fluctuations~\cite{GaGoMo,GoGaZo}. The subsequent
Sec.~4 shows the comparison of the predictions of the model with the
most recent experimental results.

\subsection{Main Assumptions}
\noindent 1. The basic assumption of the SMES is that the
production of new degrees of freedom during the early stage of A+A
collisions is a statistical process. Thus formation of all
microscopic states allowed by conservation laws is equally
probable.
As particle creation from energy does not produce net charges,
only states  with  total baryon, flavor and electric charge
quantum numbers equal to zero are considered. Presence of the
colliding nucleons is assumed to affect the properties of the
observed final state only via their interactions with the
statistically produced particles during the expansion of the
system. This issue is further discussed in point~5 below.
Consequently, the properties of the state produced at the early
stage are entirely defined by the available energy and the volume
in which production takes place. In central A+A collisions this
volume is chosen as the Lorentz contracted volume occupied by the
colliding nucleons (participant nucleons) from a single nucleus:
\begin{equation}\label{volume}
V = \frac {V_0} {\gamma}~,
\end{equation}
where $V_0 = \frac{4}{3} \pi r_0^3 A_p$ and $\gamma = \sqrt{s}_{NN}/(2
m_N)$, $m_N$ is the nucleon mass 
and $A_p$ is the number of participant
nucleons from a single nucleus. The $r_0$ parameter is taken to be
1.30 fm in order to fit the mean baryon density in the nucleus,
$\rho_0 = 0.11$ fm$^{-3}$.

\vspace{0.2cm} \noindent 2. Only a fraction, $\eta$, of the total energy in
A+A collision  is transformed into the energy of new degrees of
freedom created in the early stage. This is because a part of the
energy is carried by the  net baryon number which is conserved
during the collision. The released (inelastic) energy can be expressed as:
\begin{equation}\label{energyin}
E = \eta (\sqrt{s}_{NN} - 2 m_N)~A_p~.
\end{equation}
The parameter $\eta$ is assumed to be independent of the collision
energy and the system size for A+A collisions. The value of $\eta$
used for the numerical calculations is 0.67~\cite{Ba:94}.

\vspace{0.2cm} \noindent 3. The elementary particles of strong
interactions are quarks and gluons. The deconfined state is
considered to be composed of $u$, $d$ and $s$ quarks and the
corresponding anti-quarks each with internal number of degrees of
freedom equal to 6 (3 color states $\times$ 2 spin states).
The contribution of $c$, $b$ and $t$ quarks can be neglected due
to their  large masses.
The internal number of degrees of freedom for gluons is 16 (8
color states $\times$ 2 spin states). The masses of gluons and
non-strange (anti-)quarks are taken to be 0, the strange (anti-)quark
mass is taken to be 175 MeV \cite{Le:96}.
%
The properties of equilibrated matter is characterized by an equation of
state (EoS). For the case of colored quarks and gluons the model assumes
the ideal gas EoS  modified by a bag constant $B$
(see, e.g., \cite{qgp2,bag}):
\begin{equation}
p = p^{id} - B~,~~~
\varepsilon = \varepsilon^{id} + B~,
\end{equation}
where $p$ and
$\varepsilon$ denote pressure and energy density, respectively,
and the superscript $^{id}$ marks the quantities for the ideal gas.
This equilibrium state is called the Quark Gluon Plasma or
Q--state.

\vspace{0.2cm}
\noindent
4.
The model uses an effective parametrization of the confined state,
denoted as W--state (White--state). The non-strange
degrees of freedom which dominate the entropy production are taken
to be massless bosons, as suggested by the original analysis
of entropy production in N+N and A+A collisions \cite{Ga:95-97}.
Their internal number of degrees of freedom was fitted to the
data \cite{Ga:95-97} and appeared to be about 3 times lower
than the internal number of effective degrees of freedom in the
QGP.
The internal number of degrees of freedom for a QGP is 16 +
(7/8)$\cdot$36 $\cong$ 48 and therefore the internal number of
non-strange degrees of freedom for low energy collisions is taken
to be 48/3 = 16. The mass of strange degrees of freedom is assumed
to be 500 MeV, equal to the kaon mass. The internal number of
strange degrees of freedom is estimated to be 14 as suggested by
the fit to the strangeness and pion data at the AGS.
Also for the W-state the ideal gas EoS is selected. Clearly, this
description of the confined state should only be treated as
an effective parametrization. The numerical parameters are fixed
by fitting A+A data at the AGS and the parametrization is then used
for extrapolation to higher collision energies where the
transition between the confined and deconfined state is expected.

\vspace{0.2cm} \noindent 5. It is assumed that the matter created
at the early stage expands, hadronizes and freezes-out. Within the
original SMES formulation these later stage were not modelled. It
was, however, postulated that during these stages the total
entropy and total number of $s$ and $\overline{s}$ quarks created
in the early stage are conserved. The only process which changes
the entropy content of the produced matter during the expansion is
assumed to be the interaction with the baryonic subsystem. It was
argued that this leads to  an entropy transfer to baryons which
corresponds to the effective absorption of about 0.35
$\pi$--mesons per baryon \cite{GaGoMo:98}. Thus the final hadronic
state has non--zero baryonic number and electric charge.

\subsection{Analytical Formulas}
In this section the simplified version of
the model (massless particles) will be discussed which allows
to perform calculations analytically. Subsequently in Sec.~4
quantitative results from numerical calculations using
finite masses will be presented and compared to measured data.
All chemical potentials have to be equal to zero, as
only systems with  all conserved charges equal to zero are
considered. Thus the
temperature $T$ remains the only independent thermodynamical
variable.
It is convenient to define the EoS in terms of the pressure
function $p=p(T)$ as the entropy and energy densities can be
calculated from the thermodynamical relations:
\begin{equation}\label{termid}
s(T)~=~\frac{dp}{dT}~,~~~~\varepsilon (T)~=~T\frac{dp}{dT}~-~p~.
\end{equation}
In the case of an ideal gas the pressure of the
particle species `$j$' is given by:
\begin{equation}\label{pressi}
p^j(T)~=~\frac{g^j}{2\pi^2}\int_{0}^{\infty}k^2dk~\frac{k^2}
{3(k^2+m_j^2)^{1/2}}~\left[\exp\left(\frac{\sqrt{k^2+m_j^2}}{T}\right)
~\pm ~1\right]^{-1}~,
\end{equation}
where $g^j$ is the internal number of degrees of freedom (degeneracy
factor) for the $j$--th species, $m_j$ is the mass of the particle,
`--1' appears in Eq.~(\ref{pressi}) for bosons and `+1' for fermions.
The pressure $p(T)$ for an ideal gas of several particle
species is additive: $p(T)=\sum_j p^j(T)$. The same is
valid for the entropy and energy densities of Eq.~(\ref{termid}).

In order to be able to perform analytical calculations of the system entropy
and illustrate the model properties it is assumed
that all degrees of freedom are massless. In this simplified case
the pressure Eq.~(\ref{pressi}) is equal to:
\begin{equation}\label{presso}
p^j(T)~=~\frac{\sigma^j}{3}~T^4~,
\end{equation}
where $\sigma^j$ is the so called Stephan--Boltzmann constant,
equal to $\pi^2g^j/30$ for bosons and $\frac{7}{8}\pi^2 g^j/30$
for fermions. The total pressure in the ideal gas of several
massless species can then be written as $p(T)=\pi^2 g T^4/90$
with the effective number of degrees of freedom $g$ given  by
\begin{equation}\label{g}
g~=~g^{b}+\frac{7}{8}~g^{f},
\end{equation}
where $g^b$ and $g^f$ are internal degrees of freedom
of all bosons and fermions, respectively.
The $g$ parameter  is taken to be
$g_W$ for the W--state and
$g_Q$ for the Q--state, with $g_Q >  g_W$.

The pressure, energy and entropy densities then follow as:
\begin{equation}\label{wmatter}
p_W(T)=\frac{\pi^2g_W}{90}~T^4~,~~~
\varepsilon_W(T)=\frac{\pi^2g_W}{30}~T^4~,~~~
s_W(T)=\frac{2\pi^2g_W}{45}~T^3~,\\
\end{equation}
\begin{equation}\label{qmatter}
p_Q(T)=\frac{\pi^2g_Q}{90}~T^4-B~,~~~
\varepsilon_Q(T)=\frac{\pi^2g_Q}{30}~T^4+B~~~,~
s_Q(T)=\frac{2\pi^2g_Q}{45}~T^3~,
\end{equation}
for the pure W-- and Q--state, respectively.
Note the presence of the non--perturbative bag terms
in addition to the ideal quark--gluon gas expressions for
the pressure and energy density of the Q--state.

The 1$^{st}$ order phase transition
between W-- and Q--state is defined by the Gibbs criterion
\begin{equation}\label{ptr}
p_W(T_c)~=~p_Q(T_c)~,
\end{equation}
from which  the phase transition temperature can be calculated as:
\begin{equation}\label{tcr}
T_c~=~\left[\frac{90B}{\pi^2(g_Q-g_W)}\right]^{1/4}~.
\end{equation}
At $T=T_c$ the system is in the {\it mixed} phase with
the energy and entropy densities given by
\begin{equation}\label{mixed}
\varepsilon_{mix}=(1-\xi)\varepsilon_W^c~+~\xi
\varepsilon_Q^c~,~~~~
s_{mix}=(1-\xi)s^c_W~+~\xi s_Q^c~,
\end{equation}
where $(1-\xi)$ and $\xi$ are the relative volumes
occupied by the W-- and Q--state, respectively.
~From Eqs.~(\ref{wmatter}, \ref{qmatter}) one finds
the energy density discontinuity (`latent heat')
\begin{equation}\label{lheat}
\Delta \varepsilon ~\equiv ~ \varepsilon_Q(T_c)-
\varepsilon_W(T_c)~\equiv ~\varepsilon_Q^c-
\varepsilon_W^c~=~4B~.
\end{equation}

The early stage energy density is an increasing function of the
collision energy and is given by (see Eqs.~(\ref{volume},
\ref{energyin})):
\begin{equation}\label{endensity}
\varepsilon ~ \equiv~ \frac {E}
{V} ~=~\frac{\eta~\rho_0~(\sqrt{s}_{NN}-2m_N)~\sqrt{s}_{NN}}{2 m_N}~.
\end{equation}

\begin{figure}[!htb]
\begin{center}
\begin{minipage}[b]{1.0\linewidth}
\includegraphics[width=0.7\linewidth]{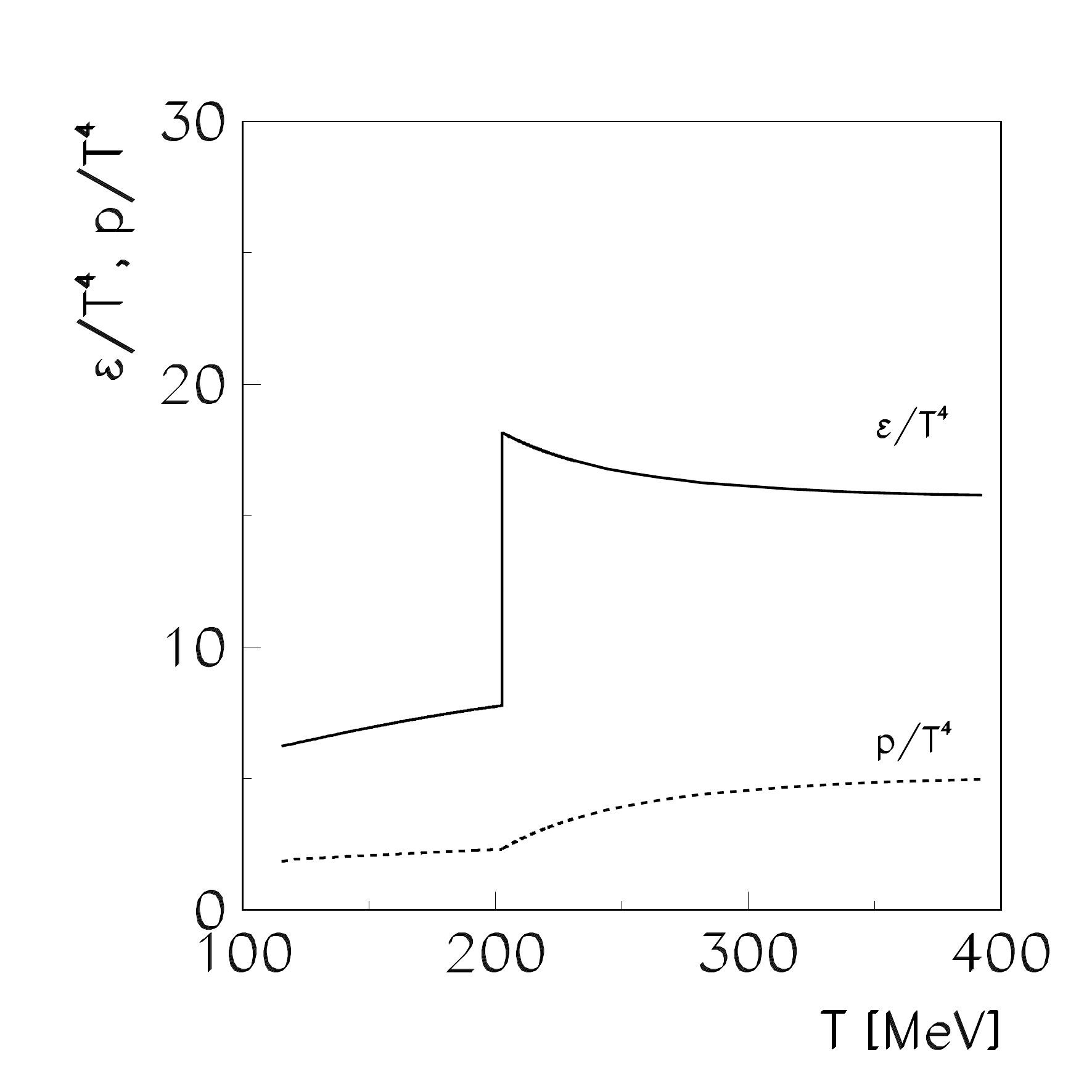}
\end{minipage}
\caption{\label{ept4}
Energy density and pressure divided by $T^4$
as a function of temperature~$T$. The bag constant $B$ was adjusted
to 600~MeV/fm$^3$ to obtain a critical temperature $T_c~=~200$~MeV.
}

\end{center}
\end{figure}

There is
a remarkable equivalence
(see Appendix C of Ref.~\cite{GaGo})
of the Gibbs criterion (i.e. the pure phase
corresponds to the larger pressure $p_W$ or $p_Q$ and the mixed phase
to equal pressures $p_W=p_Q$)  and the maximum entropy criterion,
\begin{equation}\label{extremum}
s(\varepsilon)~=~\max~\{~s_W(\varepsilon),~
s_Q(\varepsilon),~s_{mix}(\varepsilon)~\}~,
\end{equation}
for an arbitrary EoS $p=p(T)$ with a 1$^{st}$ order phase transition.
For $\varepsilon
< \varepsilon_W^c$ or $\varepsilon > \varepsilon_Q^c$ the system
consists of pure W-- or Q--state, respectively, with entropy
density given by the following equations:
\begin{equation}\label{entropyw}
s_W(\varepsilon)~=~\frac{4}{3} \left(\frac{\pi^2 g_W}{30}\right)^{1/4}~
\varepsilon^{3/4}~,
\end{equation}
\begin{equation}\label{entropyq}
s_Q(\varepsilon)~=~\frac{4}{3}\left(\frac{\pi^2 g_Q}{30}\right)^{1/4}~
(\varepsilon -B)^{3/4}~.
\end{equation}
For $\varepsilon_W^c<\varepsilon<\varepsilon_Q^c$
the system is in the mixed phase~(Eq.~\ref{mixed}) and its  entropy
density can be expressed as:
\begin{equation}\label{entropym}
s_{mix}(\varepsilon)~=~ \frac{\varepsilon_Q^cs_W^c - \varepsilon_W^cs_Q^c}
{4B}~+~\frac{s_Q^c - s_W^c}{4B}~\varepsilon~
\equiv~a~+~b~\varepsilon ~.
\end{equation}
The ratio of the total entropy
of the created state to the number of nucleons participating in A+A
collisions is:
\begin{equation}\label{Entropy}
\frac{S}{2A_p}~=~ \frac{V~s}{2A_p}~=~\frac{
m_N~s}{\rho_0\sqrt{s}_{NN}}~,
\end{equation}
and is independent of the number of participant nucleons. The
entropy density $s$ in Eq.~(\ref{Entropy}) is given by the general
expressions Eq.~(\ref{extremum}) with $\varepsilon$ defined by
Eq.~(\ref{endensity}). For small $\sqrt{s}_{NN}$ the energy
density Eq.~(\ref{endensity}) corresponds to that of the pure W--state and one
finds
\begin{equation}\label{Entropyw}
\left(\frac{S}{2A_p}\right)_{W}~=~C~g_W^{1/4}~F~,
\end{equation}
where
\begin{equation}\label{c}
C~=~\frac{2}{3}\left
(\frac{\pi^2m_N}{15\rho_0}\right)^{1/4}~\eta^{3/4}~,~~~~
F~ =~ \frac { (\sqrt{s}_{NN} - 2 m_N)^{3/4} }  { (\sqrt{s}_{NN})^{1/4} }~.
\end{equation}
Thus for low collision energies, where the W--state is created, the
entropy per participant nucleon is proportional to $F$.
For high $\sqrt{s}_{NN}$ the pure Q--state is formed
and Eq.~(\ref{Entropy}) leads to
\begin{eqnarray}\label{Entropyq}
\left(\frac{S}{2A_p}\right)_Q~&=&~C~g_Q^{1/4}~F~\left(1~-~
\frac{2m_N B}{\eta \rho_0 (\sqrt{s}_{NN}-2m_N)\sqrt{s}_{NN}}
\right)^{3/4} \\
\nonumber
& \cong&~C~g_Q^{1/4}~F~\left(1-\frac{3m_NB}{2\eta\rho_0
F^4}\right)~.
\end{eqnarray}
For large values of $F$ the entropy per participant nucleon in the Q--state
is also proportional to $F$.
The slope
is, however, larger than the corresponding slope for  the W--state
by a factor $(g_Q/g_W)^{1/4}$.
In the interval of $F$ in which the mixed phase is formed the energy
dependence of the entropy per participant nucleon is given by:
\begin{equation}\label{Entropym}
\left(\frac{S}{2A_p}\right)_{mix}~=~ \frac{C_1}{\sqrt{s}_{NN}}~+~
C_2~(\sqrt{s}_{NN}-2 m_N)~,
\end{equation}
where
\begin{equation}\label{c1}
C_1~=~\frac{m_N}{\rho_0}~a~,~~~~ C_2~=~\eta ~b~.
\end{equation}
Equation~(\ref{Entropym}) gives approximately a quadratic increase with $F$
of the entropy per participant nucleon in the mixed phase region.

\begin{figure}[!htb]
\begin{center}
\begin{minipage}[b]{1.0\linewidth}
\includegraphics[width=0.5\linewidth]{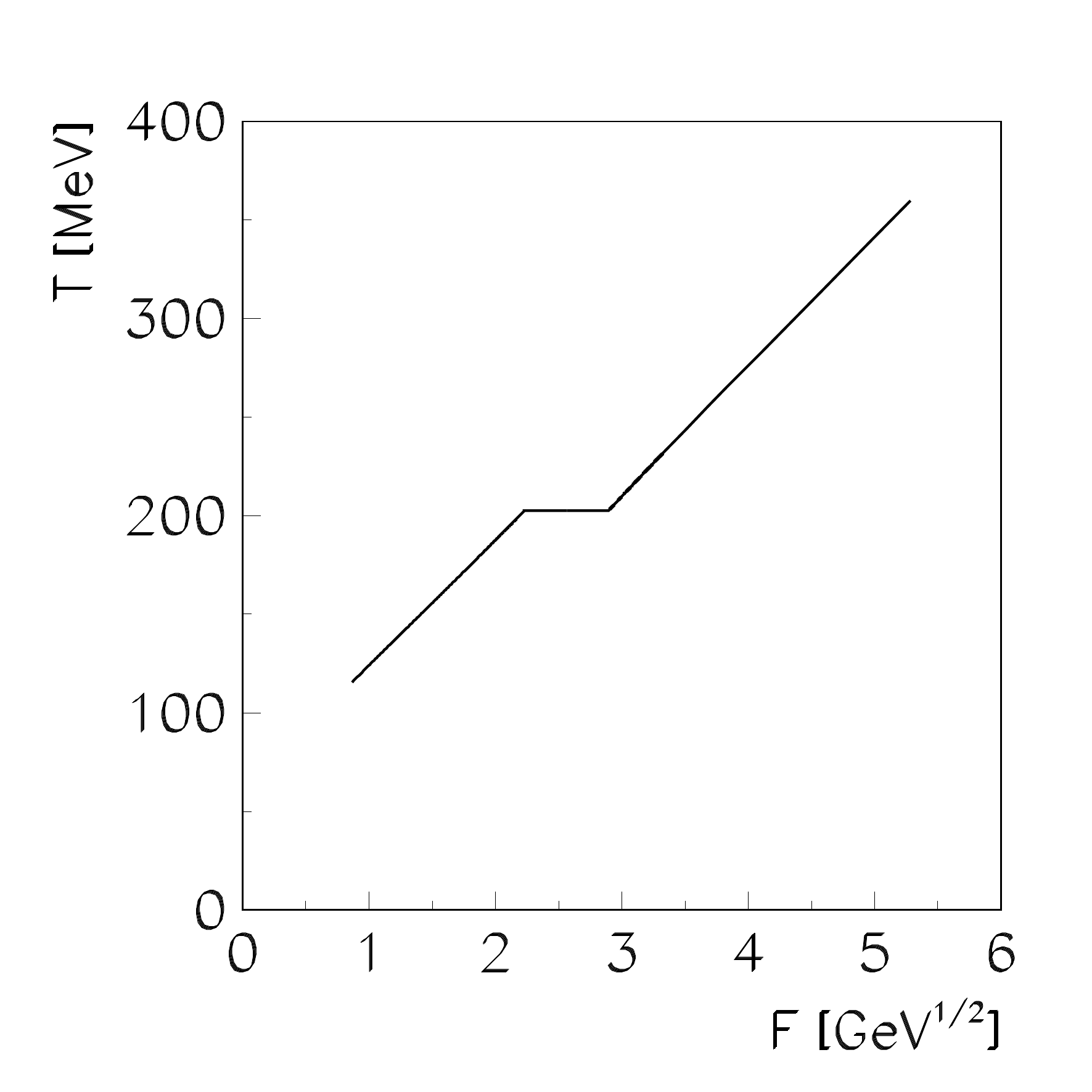}
\includegraphics[width=0.5\linewidth]{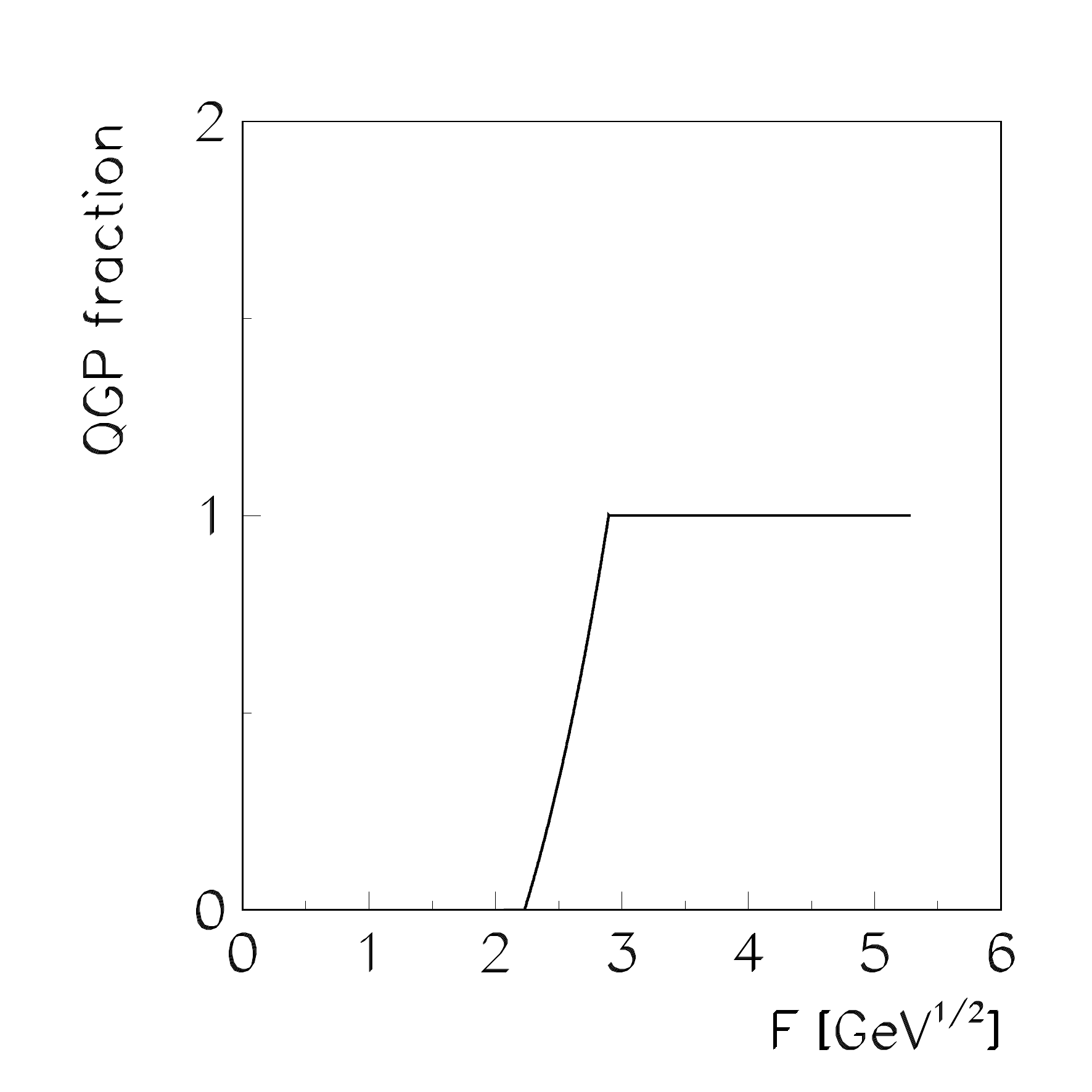}
\end{minipage}
\caption{\label{temp}
{\it Left:} The early stage (initial) temperature of the fireball as a function of
$F$. {\it Right:} The fraction of volume occupied by the QGP as a
function of $F$.
}
\end{center}
\end{figure}

Let us now turn to strangeness. The model defines  $g^s_W$ and $g^s_Q$ as
the numbers of internal degrees of freedom of (anti-)strangeness
carriers in the W-- and Q--state, respectively.
The total entropy of the considered state is given by the sum of entropies
of strange and non-strange degrees of freedom.
Provided that all particles are massless the fraction of entropy
carried by strange (and anti-strange) particles is proportional
to the number of strangeness degrees of freedom:
\begin{equation}\label{strentr}
S_s ~=~ \frac {g^s} {g}~ S~.
\end{equation}
Equation~(\ref{strentr}) is valid for both W-- and Q--state.
Note that all degeneracy factors are calculated according
to the general relation Eq.~(\ref{g}).
For massless particles of the $j$--th
species the entropy is proportional to the
particle number:
\begin{equation}
S_j ~= ~4 N_j~.
\end{equation}
Thus the number of strange and anti-strange particles can
be expressed as
\begin{equation}
N_s + N_{\overline{s}} ~=~ \frac {S}{4}~ \frac {g^s} {g}~,
\end{equation}
and the strangeness to entropy ratio is equal to
\begin{equation}\label{strent}
\frac { N_s + N_{\overline{s}} } {S}
~ =~ \frac {1}{4}~ \frac  {g^s} {g}~.
\end{equation}
One concludes therefore that the strangeness to entropy ratio for
the ideal gas of massless particles is dependent only on the ratio
of strange  to all degrees of freedom, $g^s/g$. This ratio is
expected to be equal to $g_Q^s/g_Q\cong 0.22$ in the Q--state and
$g_W^s/g_W\cong 0.5$ in the W--state.
Consequently the phase transition from the W-- to the Q--state should lead to a
decrease of the strangeness to entropy ratio by a factor of about
2. This simple picture will be modified because of the
large value of the mass of strange degrees of freedom in the W--state
($m_W^s\cong500$ MeV) compared to the temperature $T$. In this case the left
hand side of Eq.~(\ref{strent}) is a strongly increasing  function
of $T$.

In order to demonstrate properties of the EoS the ratios of
$\varepsilon/T^4$ and  $p/T^4$ are plotted in Fig.~\ref{ept4}
as  functions of the temperature. The bag constant $B$ = 600
MeV/fm$^3$ was adjusted  such that the resulting critical temperature
$T_c$ is equal to 200 MeV.

\subsection{Quantitative Calculations}

We next turn to the results from numerical calculations based on the
model using finite strangeness carrier masses.
For the number of non-strange degrees of freedom $g_Q^{ns}$ and $g_W^{ns}$
one gets as in the simplified model:
\begin{equation}\label{gnonstrange}
g_Q^{ns}~=~2\cdot8~+~\frac{7}{8}\cdot 2\cdot 2\cdot 3\cdot 2 ~=~37~;
~~~~~~g_W^{ns}~=~16~.
\end{equation}
The strange degrees of freedom are now considered to have
realistic masses $m_Q^s$ and $m_W^s$. Equation~(\ref{pressi}) is used with
\begin{equation}\label{gstrange}
g_Q^s~=~2\cdot 2\cdot 3  ~=~12~,~~m_Q^s~\cong ~ 175~ \mbox{MeV}~;
~~~~~~ g_W^s~=~14~,~~m_W^s~\cong~ 500~ \mbox{MeV}~.
\end{equation}
Note that there is no factor `7/8' in the above expression for
$g_Q^s$ as Eq.~(\ref{pressi}) with Fermi momentum distribution
was taken.
The contributions of strange degrees of freedom to
the entropy and energy densities are calculated using
the thermodynamical relations Eq.~(\ref{termid}).

A convenient variable to study collision energy dependence is the
Fermi--Landau variable $F$~ defined in Eq.~(\ref{c}).
The dependence of the early stage temperature $T$ (initial temperature
of the fireball) on
$F$ in the SMES is shown in Fig.~\ref{temp}~{\it left}. Outside the
transition region $T$ increases  approximately linearly with $F$.
Inside the transition region $T$ is constant ($T = T_c$~=~200~MeV).
The transition region begins at $F= 2.23$ GeV$^{1/2}$
($p_{LAB} = 30A$~GeV) and ends at $F= 2.90$ GeV$^{1/2}$
($p_{LAB} = 64A$~GeV). The fraction  of the volume occupied
by the Q--state, $\xi$, increases rapidly in the transition
region, as shown in Fig.~\ref{temp}~{\it right}.

\begin{figure}[!htb]
\begin{center}
\begin{minipage}[b]{1.0\linewidth}
\includegraphics[width=0.5\linewidth]{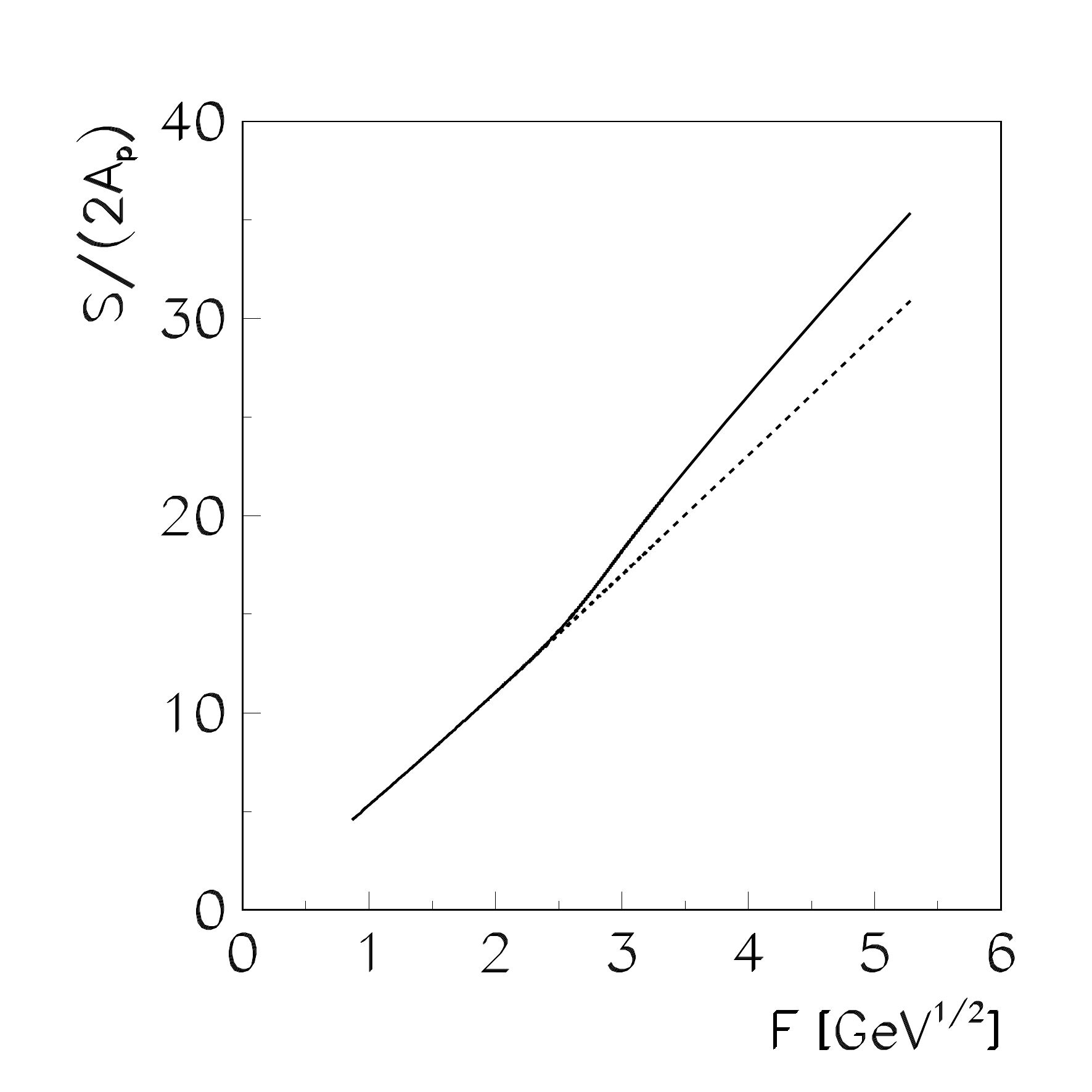}
\includegraphics[width=0.5\linewidth]{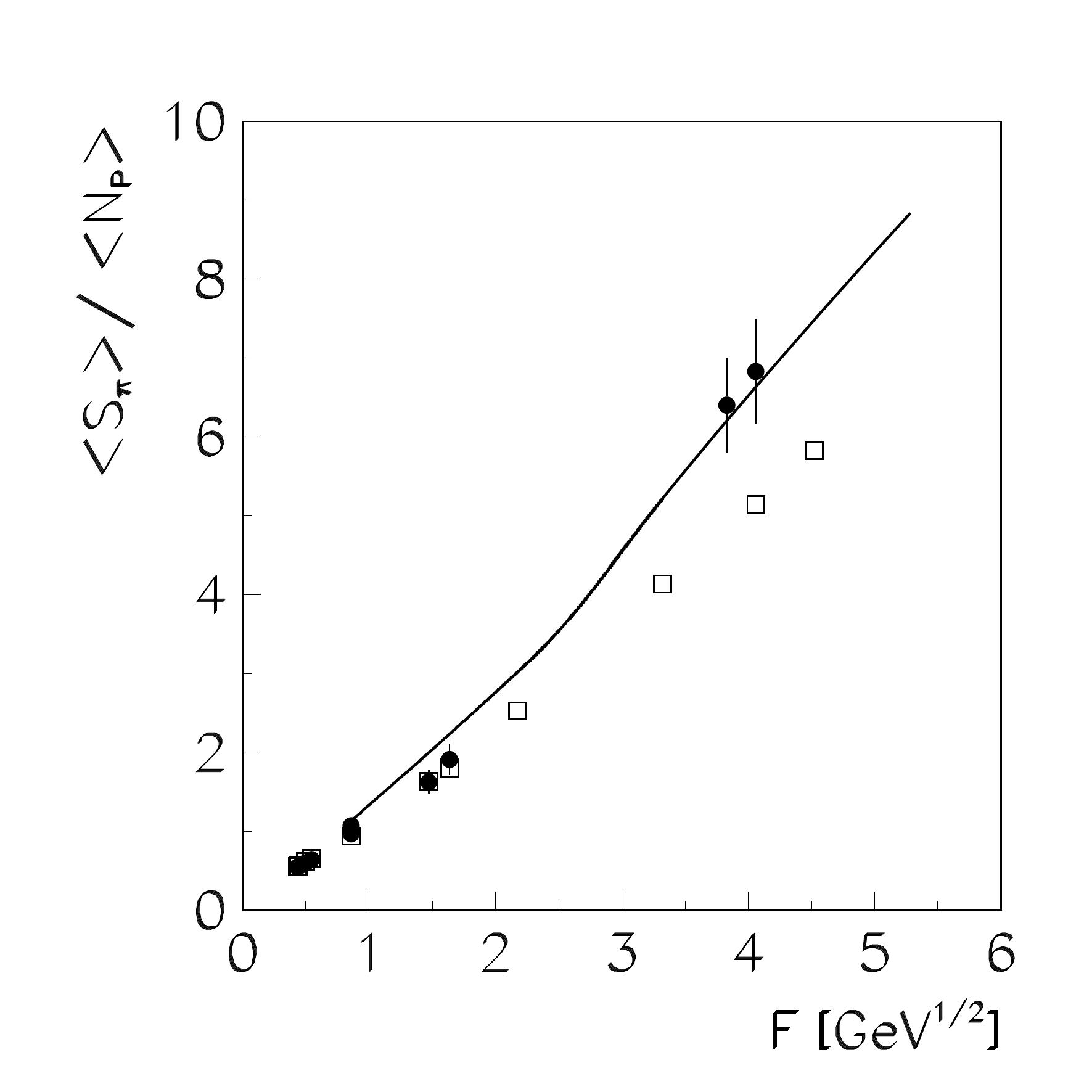}
\end{minipage}
\caption{\label{spb}
{\it Left:} The entropy per participant nucleon as a
function of $F$ (solid line). Dashed line indicates  the
dependence obtained assuming that there is no transition to the
QGP. {\it Right:} Ratio of produced entropy in pion units per
participant nucleon, $\langle S_{\pi} \rangle/\langle N_P
\rangle$, as a function of $F$. Experimental data on central
collisions of two identical nuclei are indicated by closed
circles. These data correspond to the status of 1998~\cite{GaRo1,GaRo2}
and should be compared with
the model predictions shown  by the solid line. The open boxes
show results obtained for nucleon--nucleon interactions.
}
\end{center}
\end{figure}

\begin{figure}[!htb]
\begin{center}
\begin{minipage}[b]{1.0\linewidth}
\includegraphics[width=0.4\linewidth]{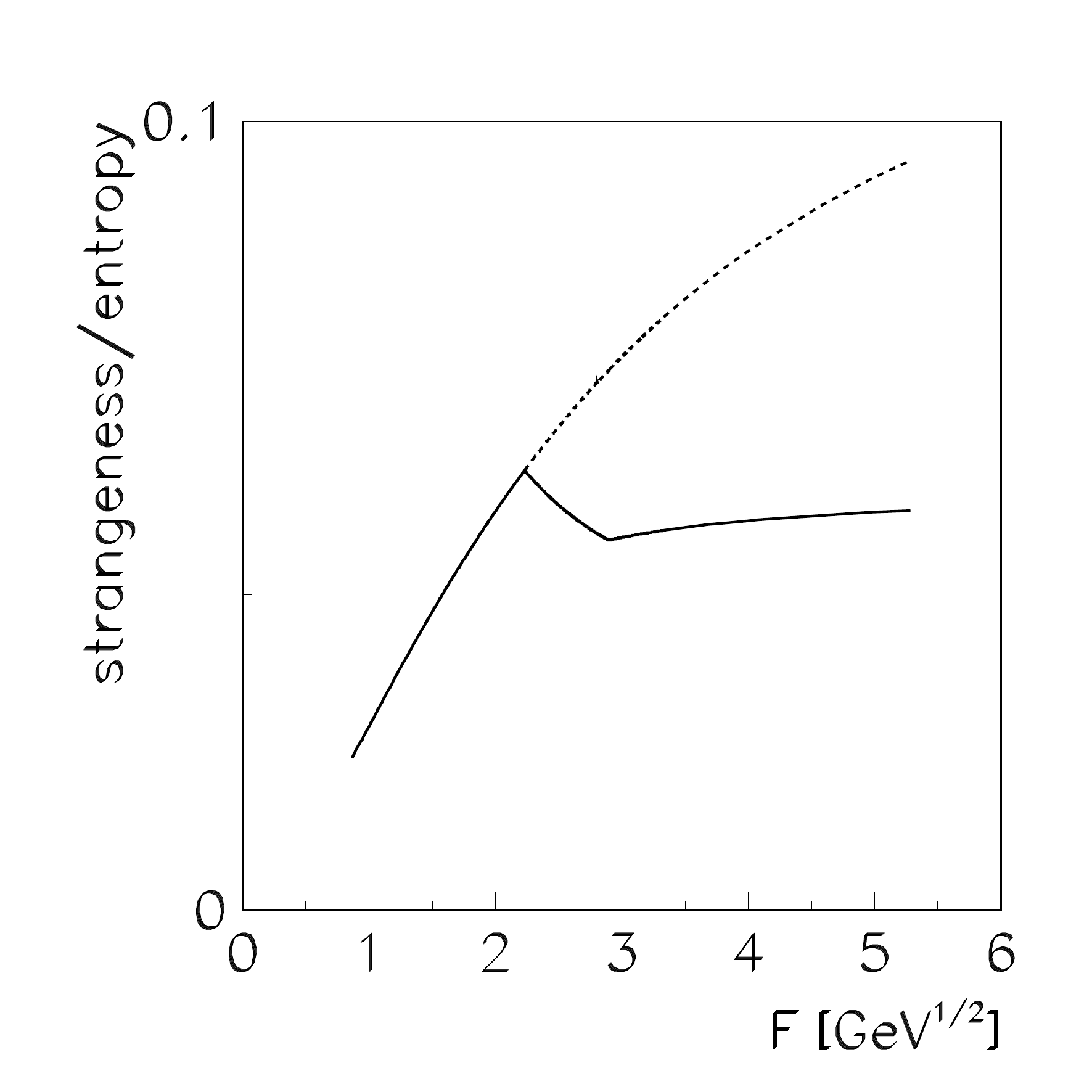}
\includegraphics[width=0.4\linewidth]{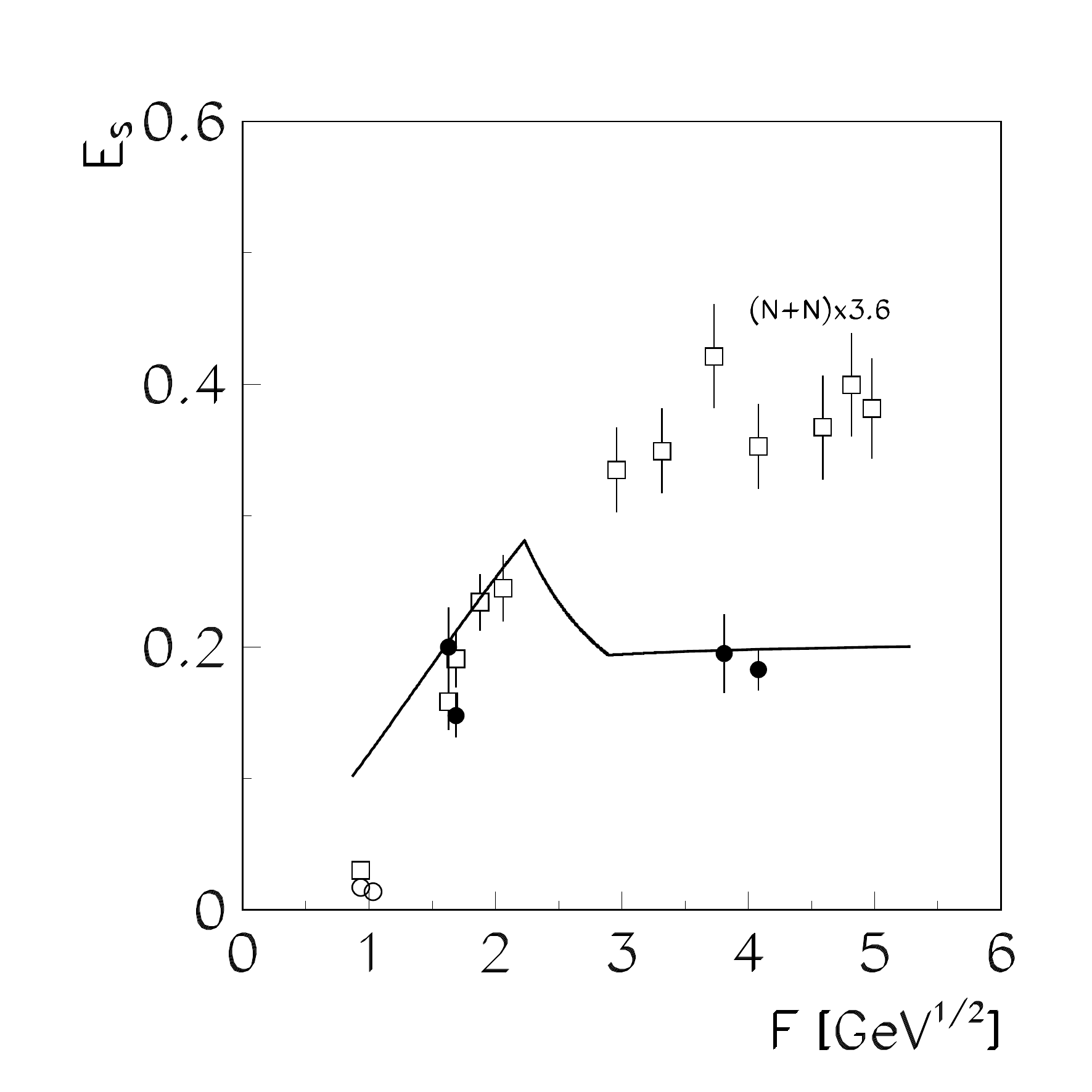}
\end{minipage}
\caption{\label{str}
{\it Left:} The ratio of the total number of $s$ and
$\overline{s}$ quarks and anti-quarks to the entropy (solid line)
as a function of $F$. The dashed line indicates the corresponding
ratio calculated assuming absence of the phase transition to the
QGP. {\it Right:} The ratio $E_S$ of strange particle to
pion production as a function of $F$. Experimental
data on central collisions of two identical nuclei are plotted
as closed circles. These data should be compared with the model
predictions shown  by the solid line. The open boxes show results
obtained for nucleon--nucleon interactions, scaled be a factor 3.6
to match A+A data at AGS energy. The plotted data show the status of
1998 as compiled in~\cite{GaRo1,GaRo2}.
}
\end{center}
\end{figure}

The number of non-strange and strange  degrees of freedom and their
masses are given by Eqs.~(\ref{gnonstrange}, \ref{gstrange}). They
are different in the confined and deconfined phases. Thus, one
expects abrupt changes of the pion multiplicity (entropy) (see
Fig.~\ref{spb}~{\it left}) and the
multiplicity of strange particles (see Fig.~\ref{str}~{\it left})
as a function of collision
energy in the energy range where a transition from confined to
deconfined matter takes place at the early stage of A+A collisions.
The comparison of these predictions with experimental results
is discussed in Secs.~4.1 and~4.2.

\subsection{Extensions of the SMES}

Several extensions of the SMES were developed over the past 10 years
which allow to predict
signals of the onset of deconfinement related to the
collective flow at freeze--out~\cite{GoGaBu,bleicher} and to
event-by-event fluctuations~\cite{GaGoMo,GoGaZo}.
These extensions are briefly presented in this subsection.

\subsubsection{Collective Flow at Freeze-out}

The collective flow of matter at  freeze-out depends on the
properties of the early stage as well as on the expansion dynamics
itself. Within the SMES the collision energy dependence of the
early stage properties is predicted. In particular, in the energy
range in which the mixed phase is created the pressure and
temperature are constant and at the end of the mixed phase domain
the pressure to energy density ratio reaches its minimum (the
softest point of the EoS). From general hydrodynamic
considerations this is expected to lead to a reduction of the
buildup of transverse~\cite{GoGaBu} and
longitudinal~\cite{bleicher} collective flow at freeze-out. The
corresponding signals are discussed in Secs.~4.3 and~4.4.

\subsubsection{
Event-by-Event Fluctuations
}

Up to this point only quantities averaged over many collisions
(events) were considered. Next an extension of the SMES is
reviewed which leads to predictions of fluctuations from event to
event.

The key additional assumption is that 
when the collision energy is fixed, 
the energy, which is used for particle production (inelastic energy) can
still fluctuate.
These
dynamical energy fluctuations
lead to dynamical fluctuations of macroscopic
properties $X$ of the  matter, like its entropy and strangeness
content~\cite{GaGoMo}. The relation between them is
given by the EoS.
For example, different values of the energy of the
early equilibrium state lead to different, but uniquely
determined, entropies. 
Since the EoS shows
an anomalous behavior in the phase transition region, this anomaly
should also be visible in the ratio of entropy to energy
fluctuations~\cite{GaGoMo}.

\begin{figure}[!htb]
\begin{center}
\begin{minipage}[b]{1.0\linewidth}
\includegraphics[width=0.5\linewidth]{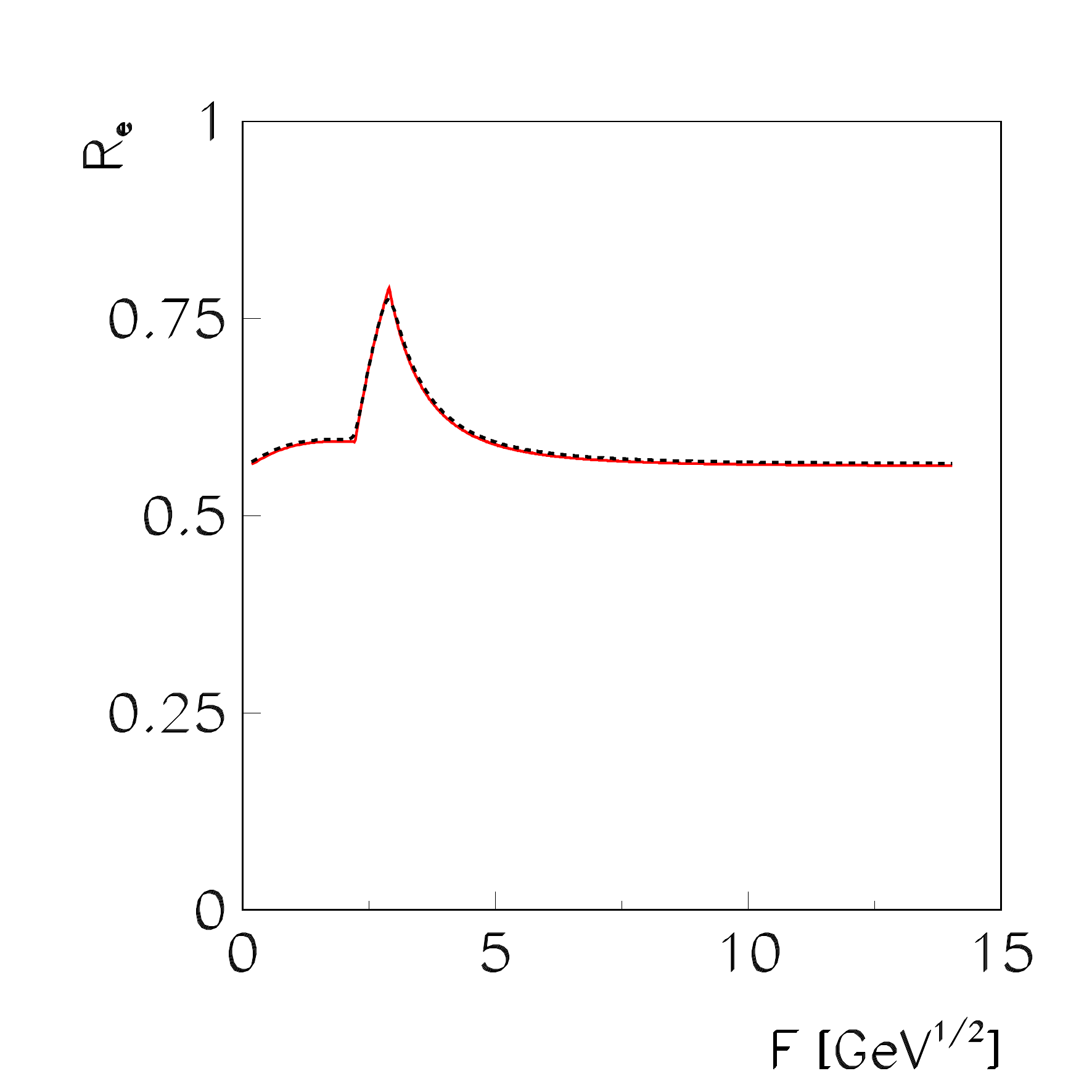}
\includegraphics[width=0.5\linewidth]{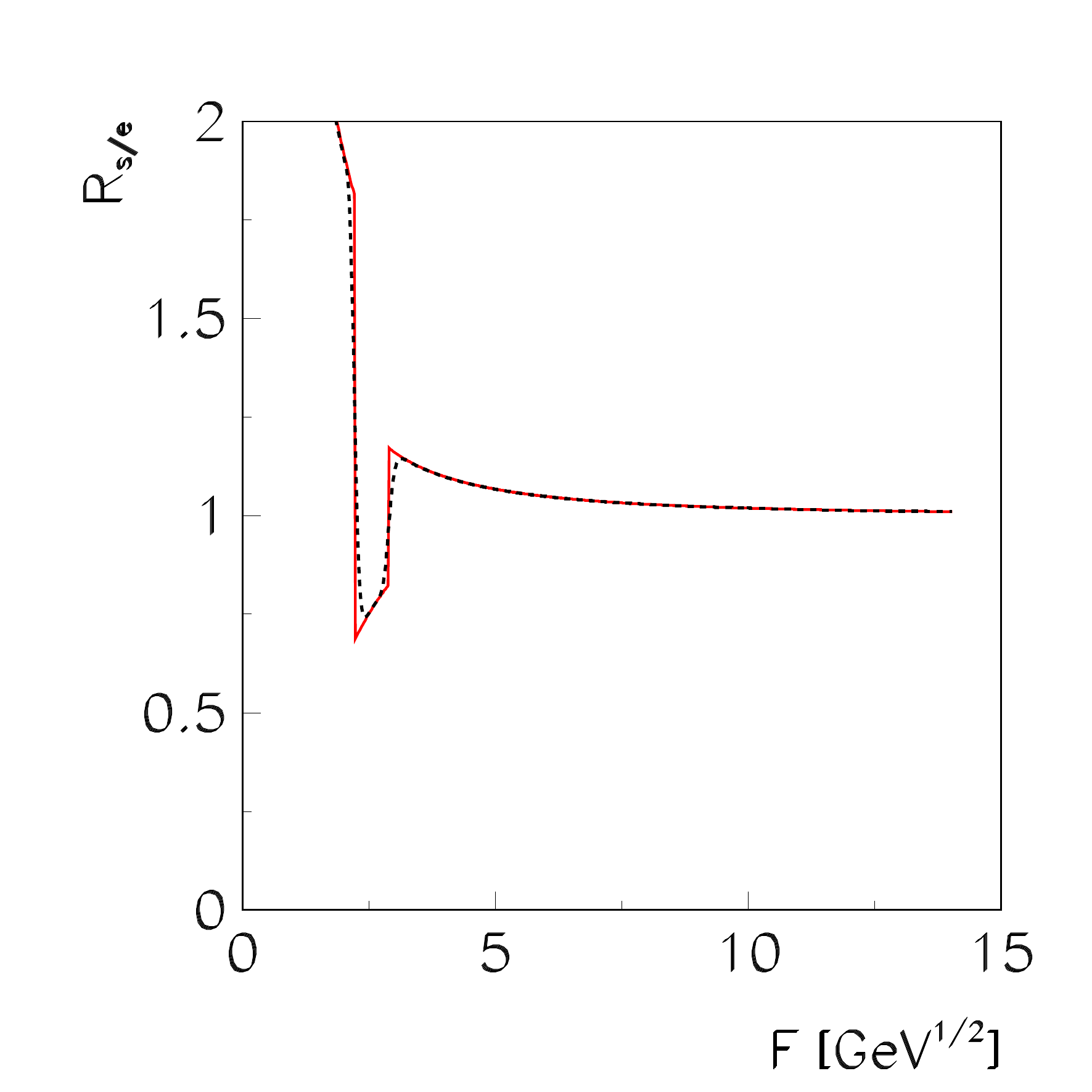}
\end{minipage}
\caption{\label{rf} The collision energy dependence of the
fluctuation signals of the onset of the deconfinement calculated
within the SMES. {\it Left:} The {\it shark fin} in  of the ratio
of entropy to energy fluctuations $R_e$ (\ref{R}) (see
Ref.~\cite{GaGoMo}). {\it Right:} The {\it tooth} structure in
the ratio of strangeness and entropy fluctuations $R_{s/e}$ (\ref{Rse}) (see
Ref.~\cite{GoGaZo}). }
\end{center}
\end{figure}

According to the first and the second principles of thermodynamics the
entropy change $\delta S$ is given as $T\delta S = \delta E + p \delta V$.
For central A+A collisions, one expects $ \delta V\cong 0$.
Within the SMES  the ratio of entropy to
energy fluctuations can then be calculated and expressed as
a simple function of the $p/\varepsilon$
 ratio \cite{GaGoMo}:
\begin{equation}\label{R}
R_e ~\equiv ~\frac{(\delta S)^2/S^2}{(\delta E)^2/E^2}~=~
\left(1~+~\frac{p}{\varepsilon}\right)^{-2}~.
\end{equation}
Within the SMES model, confined matter (which is modelled as an
ideal gas) is created at the early collision stage below a
collision energy of 30$A$~GeV. In this domain, the ratio
$p/\varepsilon$, and consequently the $R_e$ ratio, are
approximately independent of the collision energy and equal about
1/3 and 0.56, respectively. The SMES model assumes that the
deconfinement phase transition is of the first order. Thus, there
is a mixed phase region, corresponding to the energy interval
30$A$--60$A$~GeV. At the end of this region the
$p/\varepsilon$ ratio reaches a minimum (the `softest point' of
the EoS \cite{Hu:95}). Thus in the transition energy range the
$R_e$ ratio increases and reaches its maximum, $R_{e}\approx 0.8$,
at the end of the transition domain. Further on, in the pure
deconfined phase, which is represented by an ideal quark-gluon gas
under bag pressure, the $p/\varepsilon$ ratio increases and again
approaches its asymptotic value 1/3 at the highest SPS energy of
160$A$~GeV. The numerically calculated predictions of the SMES are
plotted in Fig.~\ref{rf}~{\it left}.
The early stage energy and entropy fluctuations entering Eq.~(\ref{R})
are not directly observable, however, as argued in Ref.~\cite{GaGoMo},
they can be inferred from the experimentally accessible information
on the final state energy and multiplicity fluctuations.

In Ref.~\cite{GoGaZo} the energy dependence of dynamical
strangeness fluctuations caused by dynamical energy fluctuations
was studied within the SMES model. Defining $\overline{N}_s$ as
the total number of strange quark--anti-quark pairs created in an
A+A collision one calculates the  fluctuation ratio as:
\begin{equation}
R_s~=~\frac{(\delta \overline{N}_s)^2/\overline{N}_s^2}{(\delta
E)^2/E^2}~.
\label{Rs}
\end{equation}
For $T\rightarrow \infty$ the system is in the QGP phase. Strange
(anti-)quarks can be considered as massless and the bag constant
can be neglected. Then $\varepsilon \propto T^4$ and $n_s\propto
T^3$ and consequently $d\varepsilon /\varepsilon = 4 \cdot dT/T$
and $dn_s/n_s=3\cdot dT/T$, which results in $R_s=(3/4)^2\cong
0.56$. In the confined phase, $T<T_c$, the energy density is still
approximately proportional to $T^4$ due to the dominant
contributions of non-strange hadron constituents. However, the
dependence of the strangeness density on $T$ is governed
by the exponential factor, $n_s\propto \exp(-m_S)$, as
$T<<m_S = m_W^s \cong 500$~MeV. Therefore, at small $T$ one finds
$d\varepsilon /\varepsilon = 4\cdot dT/T$ and $dn_s/n_s = m_S\cdot
dT/T^2$, so that the ratio $R_s=m_S/(4T)$ decreases with $T$. The
strangeness density $n_s$ is small and goes to zero at
$T\rightarrow 0$, but the fluctuation ratio $R_s$ Eq.~(\ref{Rs}) is
large and increases to infinity in the zero temperature limit.
One finds a non-monotonic energy dependence of $R_e$ with a maximum
at the boundary between the mixed phase and the QGP \cite{GaGoMo}.
A pronounced minimum-structure is expected in the dependence of
$R_s$ on the collision energy \cite{GoGaZo}. It is located at
$\sqrt{s}_{NN}~=$~7--12~GeV ($30A-60A$~GeV),
where the mixed phase is created at the early stage of A+A collision.

Both entropy and strangeness fluctuation measures, $R_e$ and
$R_s$, show anomalous behavior in the transition region:
a maximum is expected for $R_e$ and a minimum for $R_s$.
Consequently, an even stronger anomaly is predicted for the ratio:
\begin{equation}
R_{s/e}~\equiv~\frac{R_s}{R_e}~=\frac{(\delta
\overline{N}_s)^2/\overline{N}_s^2}{(\delta
\overline{N}_-)^2/\overline{N}_-^2}~, \label{Rse}
\end{equation}
shown in Fig.~\ref{rf}~{\it right}. Experimental measurements of
$R_{s/e}$ may be easier than the measurements of $R_e$ and $R_s$
because the ratio $R_{s/e}$ requires measurements of particle
multiplicities only, whereas both $R_e$ and $R_s$ involve also
measurements of particle energies.

These predictions are discussed in Sec.~4.5 in the context
of existing experimental data.

\section{ Signals of the Onset of Deconfinement}

Next the predictions of the SMES model
reviewed in Sec.~3 will be related to directly measurable qunatities
and compared with 
available experimental results. In particular,
their significance as evidence for the onset of deconfinement will
be discussed in detail.

\subsection{The Kink}

The majority of all particles produced in high energy interactions
are pions. Thus, pions carry basic information on the entropy
created in the collisions. On the other hand, entropy production
should depend on the form of matter present at the early stage of
collisions. Deconfined matter is expected to lead to a final state
with higher entropy than that created by confined matter.
Consequently, it is natural to expect that the onset of creation
of deconfined matter should be signaled by an enhancement of
entropy and thus pion production. This simple intuitive argument
can be quantified within the SMES.

Equilibration between newly created matter and baryons is assumed
to take place during the evolution of the system. It was argued that this
equilibration causes transfer of entropy from the produced matter
to baryons. The analysis of the pion suppression effect at low
collision energies indicates that this transfer corresponds to the
effective absorption of about 0.35 pion per participant nucleon
\cite{GaGoMo:98}. It is further assumed that there are no other processes which
change the entropy content of the state produced during the early
stage.

For the comparison with the model it is convenient  to define
the quantity:
\begin{equation}
\langle S_{\pi} \rangle = \langle \pi \rangle + \kappa
\langle K + \overline{K} \rangle + \alpha \langle N_P \rangle~,
\end{equation}
where $\langle \pi \rangle$ is the measured total multiplicity of
final state pions and $\langle K + \overline{K} \rangle$
is the multiplicity of kaons and anti-kaons.
The  factor $\kappa$ = 1.6 is the approximate
ratio between mean entropy carried by a single kaon to the
corresponding pion entropy at chemical freeze--out. The term
$\alpha \langle N_P \rangle$ with $\alpha$ = 0.35 is the correction
for the discussed partial transfer of the entropy to baryons. The
quantity $\langle S_{\pi} \rangle$ can thus be  interpreted as the
early stage entropy measured in  pion entropy units. The
conversion factor between $S$ and $\langle S_{\pi} \rangle$ is
chosen to be 4 ($\approx$ entropy units per pion at chemical freeze--out).

The dependence of the entropy per participant nucleon  on $F$ is
shown in Fig.~\ref{spb}~{\it left}. Outside the transition region
the entropy increases  approximately proportional to $F$, but the
slope in the Q--state region is larger than the slope in the
W--state region.

The number of baryons which take part in the collision ($2A_p$ in the model
calculations) is identified now with the experimentally measured number of
participant nucleons, $\langle N_P \rangle$.
The fraction of energy carried by the produced particles
is taken to be $\eta= 0.67$~\cite{Ba:94} and is assumed
to be independent of the size
of the colliding nuclei and the collision energy.

The comparison made in 1998 between the ratio $\langle S_{\pi}
\rangle/\langle N_P \rangle$ calculated from available
measurements and the model is shown in Fig.~\ref{spb}~{\it right}.
The parametrization of the W--state was
chosen to fit the AGS data and, therefore, the agreement with
low energy A+A data is not surprising. On the other hand the
description of high energy (SPS) results obtained by the NA35 and
NA49 Collaborations is essentially parameter free.

\begin{figure}[!htb]
\begin{center}
\begin{minipage}[b]{0.8\linewidth}
\includegraphics[width=1.0\linewidth]{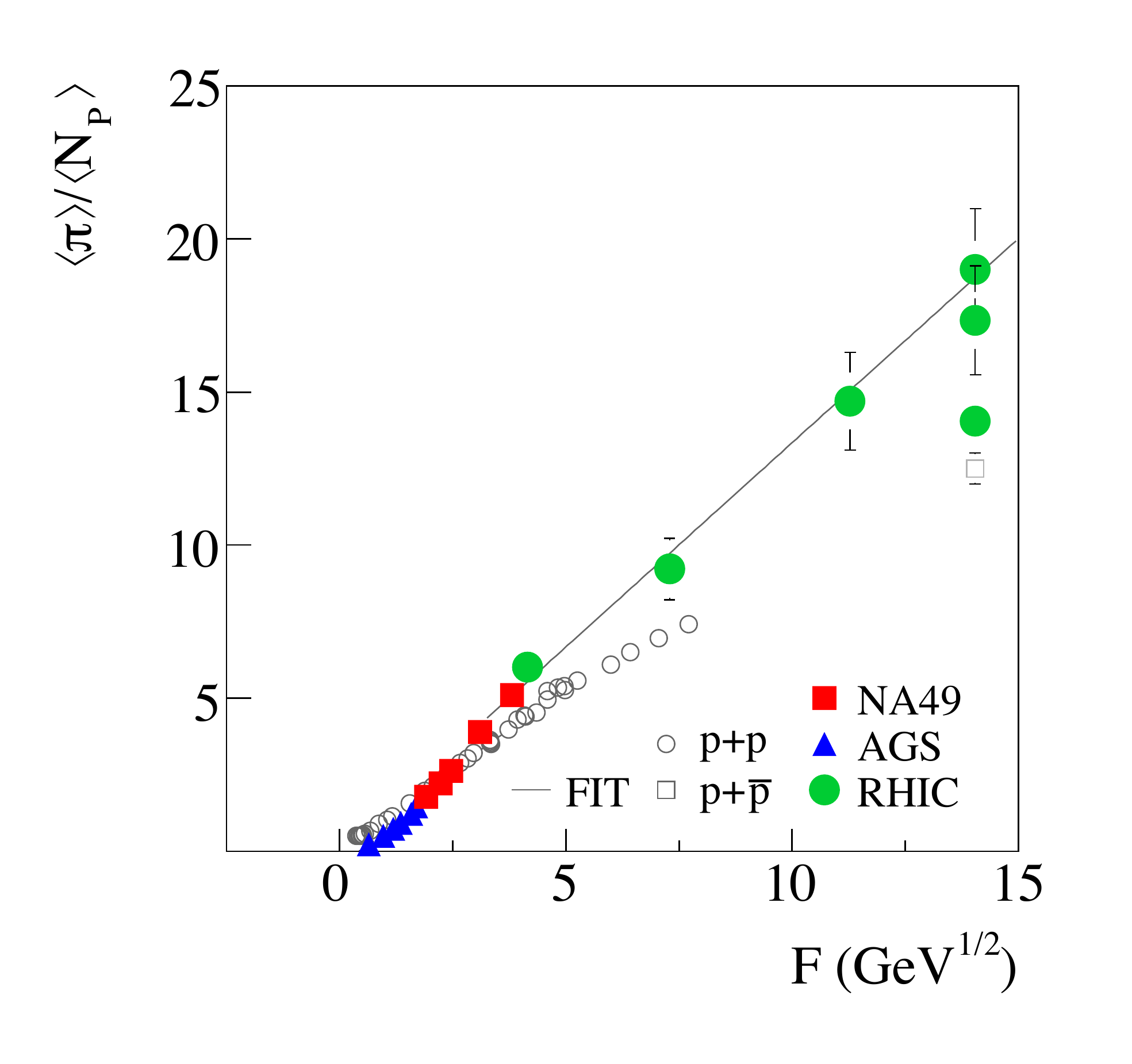}
\end{minipage}
\caption{\label{pions}
  Energy dependence of the mean pion multiplicity per participant
  nucleon measured in central Pb+Pb and Au+Au
  collisions (full symbols), compared to the corresponding results
  from $p+p(\bar{p})$ reactions (open symbols).
  The compilation of data is from
  Ref.~\cite{evidence}.
}
\end{center}
\end{figure}

The pion multiplicity is proportional to the initial
entropy, and the $\langle \pi \rangle/\langle N_P \rangle$ ratio
can thus be calculated outside the transition region as,
\begin{equation}\label{npi}
\frac{\langle \pi \rangle}{\langle N_P \rangle}~\propto~
g^{1/4}~F~,
\end{equation}
where $g=g_W^{ns}=16$ for the initial state in the confined phase
and $g=g_Q=47.5$ for the initial state in the deconfined phase
at $T >> m^s_Q$.
Therefore, the $\langle \pi \rangle/\langle N_P \rangle$ ratio
increases linearly with Fermi's energy measure $F$ outside the
transition region, and the slope parameter is proportional to
$g^{1/4}$ \cite{Ga:95-97}. In the transition region, a steepening
of the increase of pion production with energy is predicted,
because of the activation of the partonic degrees of freedom.

The recent compilation of data \cite{evidence} on
pion multiplicity produced in central Pb+Pb (Au+Au) collisions and
$p+p(\bar{p})$ interactions is shown in Fig.~\ref{pions} which displays
the mean pion multiplicity $\langle \pi\rangle  = 1.5\, (\langle
 \pi^-\rangle + \langle \pi^-\rangle )$ per wounded nucleon as a function of $F$.
The results from $p+p(\bar{p})$ interactions are shown by the open
symbols.
Up to the top SPS energy the mean pion multiplicity in $p+p$
interactions is approximately proportional to $F$.  A fit of
$\langle \pi \rangle/\langle N_P\rangle =b \cdot F$
yields a value of $b \cong 1.063$~GeV$^{-1/2}$.
For central Pb+Pb and Au+Au collisions (filled symbols in Fig.~\ref{pions})
the energy dependence is more complicated.
Below 40$A$~GeV ($\sqrt{s}_{NN}=$~8.7~GeV) the ratio $\langle \pi \rangle/\langle
N_P\rangle $ is lower in A+A collisions than in $p+p(\bar{p})$
interactions (pion suppression) while at higher energies this
ratio is larger in A+A collisions than in $p+p(\bar{p})$
interactions (pion enhancement).
A linear fit, $\langle \pi \rangle/\langle N_W\rangle =a+b \cdot F$ for
$F < 1.85$~GeV$^{1/2}$ gives $a \cong -0.45$ and $b \cong
1.03$~GeV$^{-1/2}$. The slope parameter fitted in the range $F
> 3.5$~GeV$^{1/2}$ is $b \cong 1.33$. This is shown by the
solid line in Fig.~\ref{pions} (the lowest data point at the top RHIC
energy was excluded from the fit). Thus, in the region
15$A$--40$A$~GeV between the highest AGS and the lowest SPS
energy the slope increases by a factor of about 1.3.
This agrees with the SMES result:
\eq{
\label{gg}
\left(\frac{g_Q}{g_W^{ns}}\right)^{1/4}~=~\left(\frac{47.5}{16}\right)^{1/4}
~\cong~1.31~,
}
where the increase is caused by the creation of a transient state of
deconfined matter in the early stage of the collisions at energies higher
than $\sqrt{s}_{NN}~\approx$~7.6~GeV (30$A$~GeV).

\subsection{The Horn}

The enhanced production of strangeness was considered by many
authors as a potential signal of QGP formation \cite{Rafelski, Ka:86,
Ma:86}. The line of arguments is the following. One estimates that
the strangeness equilibration time in the QGP is comparable to the
duration of the collision process ($< 10 $ fm/c) and about 10
times shorter than the corresponding equilibration time in
hadron matter. It is further assumed that in the early stage
of the fireball the
strangeness density is much below the equilibrium density, e.g., it
is given by the strangeness obtained from the superposition of
nucleon--nucleon interactions. Thus it follows that during the
expansion of the matter the strangeness content increases rapidly
and approaches its equilibrium value provided  matter is in the
QGP state. In hadron matter the modification of the
initial strangeness content is less significant due to the long
equilibration time. This leads to the expectation that strangeness
production should rapidly increase when the energy threshold for 
the production of deconfined matter is crossed from below.

In the SMES the role of strangeness is different. This is because
statistical production of particles is postulated and therefore
also strange particles are assumed to be produced in equilibrium.
Consequently possible secondary processes do not modify its value.
At $T = T_c$ the strangeness density is lower in the QGP than in
confined matter. Thus, a suppression of strangeness production is
expected to occur when crossing the transition energy range from
below.
%
The low level of strangeness production in N+N interactions as
compared to the higher strangeness yield per participant nucleon
in central A+A collisions (called strangeness enhancement) can be
understood as mostly due to the effect of strict strangeness
conservation (canonical suppression) imposed on the strange and
anti-strange degrees of freedom~\cite{danos}. This constraint has
an important effect for small statistical systems such as the
confined matter in the early stage of N+N collisions.

We are interested in the collision energy region between the AGS
and SPS. At `low' collision energies (when a pure W--state is
formed) the strangeness to entropy ratio increases with $F$. This
is due to the fact that the mass of the strange degrees of freedom
is significantly higher than the system temperature. At $T = T_c$
the ratio is higher in the W--state than in the  Q--state region.
Therefore the ratio decreases in the mixed phase region to the level
characteristic for the Q--state. In the Q--state, due to the low
mass of strange quarks compared to the system temperature,
only a weak dependence of the ratio on $F$ is observed. The $F$
dependence of the strangeness/entropy ratio calculated in
the SMES is shown in Fig.~\ref{str}~{\it left}.

The comparison between the model and the data on strangeness
production is performed under the assumption that the strangeness content
created during the early stage is preserved till the hadronic freeze--out.
This simplifies the model calculations by neglecting possible
gluon contribution to strangeness production
during hadronization of the QGP.

Total strangeness production can be studied (in the AGS and SPS
energy range) using the experimental ratio:
\begin{equation}\label{esexp}
E_S ~=~ \frac {\langle \Lambda \rangle + \langle K + \overline{K} \rangle}
            {\langle \pi \rangle}~,
\end{equation}
where $\langle \Lambda \rangle$ is the mean multiplicity of $\Lambda$
hyperons (see Appendix~8.2).
Within the SMES model $E_S$ of Eq.~(\ref{esexp}) is calculated as:
\begin{equation}\label{esmodel}
E_S ~=~ \frac {(N_s + N_{\overline{s}})/\zeta}
            { (S - S_s)/4 - \alpha \langle N_P \rangle }~,
\end{equation}
where  $\zeta$ = 1.36 is the experimentally estimated ratio
between total strangeness production and strangeness carried by  $\Lambda$
hyperons and $K$ + $\overline{K}$ mesons \cite{Bi:92} and $S_s$ is
the fraction of the entropy carried by the strangeness carriers.
The comparison between the calculations and the data available in 1998 is
shown in Fig.~\ref{str}~{\it right}~\cite{GaGo}. The good description of the AGS
data is again a consequence of the parametrization of the W-state:
$g_W^s = 14$, $m_W^s = 500$ MeV which was based on these data.
As in the case of the pion
multiplicity, the description of the strangeness results at the
SPS (from the NA35 and NA49 Collaborations) can be considered as being
essentially parameter free \footnote{ The $E_S$ value resulting
from a QGP can be estimated in a simple way. Assuming that $m_s =
0$, and neglecting the small ($< 5 \%$) effect of pion absorption
at the SPS, one gets from  Eq.~(\ref{strent}) and Eq.~(\ref{esmodel}) $E_S
\approx (g^s_Q/1.36)/g^{ns}_Q \approx 0.21$, where $g^s_Q = (7/8)
\cdot 12$ is the effective number of degrees of freedom of $s$ and
$\overline{s}$ quarks and $g^{ns}_Q = 16 + (7/8) \cdot 24$ is the
corresponding number for $u, \overline{u}, d, \overline{d}$ quarks
and gluons. Moreover, we use  the approximation that the pion
entropy at  freeze--out is equal to the mean entropy of $q$,
$\overline{q}$ and $g$ in a QGP.}. The agreement with the SPS data
is obtained assuming creation of globally equilibrated QGP in the
early stage of A+A collisions. The characteristic
non--monotonic energy dependence of the $E_S$ ratio is a signature
of the phase transition.

The entropy and strangeness production in central A+A collisions
satisfies well the conditions needed for thermodynamical treatment.
Therefore one expects that the measures of the entropy per participant
nucleon,
$\langle S_{\pi} \rangle/\langle N_P \rangle$, and
the ratio of strangeness to entropy production, $E_S$,
are independent of the number of participants
for large enough values of $\langle N_P \rangle$.

The energy dependence of the strangeness to entropy production ratio is a
crucial  signal of deconfinement.
The temperature dependence of the multiplicity of a particle is strongly
dependent on its mass. In the Boltzmann approximation one finds:
\begin{equation}\label{N}
\langle N_i \rangle ~=~ \frac{g^iV}{2 \pi^2}~ \int
\limits_{0}^{\infty} p^2 dp
~\exp\left(-~\frac{\sqrt{p^2+m_i^2}}{T}\right)~ =~ \frac{g_iV}{2
\pi^2}~ m_i^2 T~ K_{2}\left(\frac{m_i}{T} \right)~,
\end{equation}
where
$K_2$ is the modified Hankel function. For
light particles ($m_l/T\ll 1$) one finds from Eq.~(\ref{N}),
$\langle N_l \rangle \propto T^3$, whereas for heavy particles
($m_h/T\gg 1$) Eq.~(\ref{N}) leads to $\langle N_h \rangle \propto
T^{3/2}\exp(-~m_h/T)$. Within the SMES the strangeness to entropy production ratio
increases steeply at low collision energies,
when confined matter is produced. This is due to the low
temperature at the early stage and the high mass of the carriers
of strangeness (the kaon mass). Thus, $m_K\gg T$ and total
strangeness production is proportional to
$T^{3/2} \exp(-~m_K/T)$. On the other hand, the total entropy is
approximately proportional to $T^3$. Therefore, the strangeness to
pion production ratio is approximately $T^{-3/2} \exp(-m_K/T)$ in
the confined phase and strongly increases with the collision
energy.  When the transition to deconfined matter occurs, the mass
of the strangeness carriers is significantly reduced ($m_s \cong
175$~MeV, the strange quark mass). Due to the lower mass ($m_s <
T$) the strangeness yield becomes approximately proportional to
the entropy (both are proportional to $T^3$), and the strangeness
to entropy (or pion) production ratio becomes independent of
energy in the QGP. This leads to a `jump' in the energy dependence
from the larger value for confined matter to the value for
deconfined matter. Thus, within the SMES, the non-monotonic energy
dependence of the strangeness to entropy production ratio followed
by a plateau at the deconfined value is a direct consequence of
the onset of deconfinement taking place at
$\sqrt{s}_{NN}~\approx~$~7.6~GeV (about 30$A$~GeV)~\cite{GaGo}.

\begin{figure}[!htb]
\begin{center}
\begin{minipage}[b]{1.0\linewidth}
\includegraphics[width=0.5\linewidth]{kpip_4pi_data.pdf}
\includegraphics[width=0.5\linewidth]{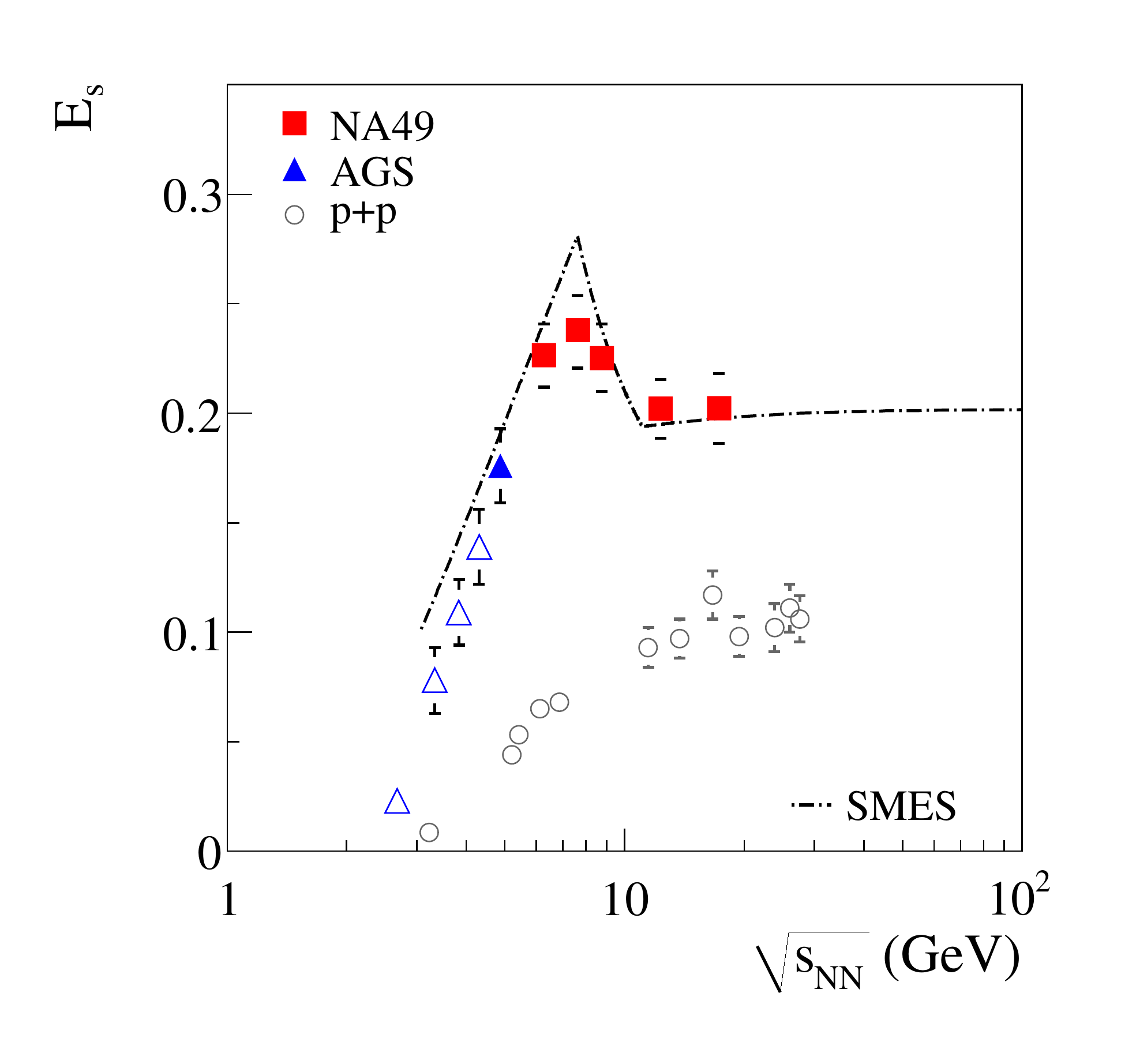}
\end{minipage}
\caption{\label{strangeness}
  {\it Left:}
    Energy dependence of the $\langle K^+ \rangle /\langle \pi^+\rangle$ ratio
    measured in central Pb+Pb and Au+Au collisions
    (upper set of symbols) compared to the corresponding results from $p+p$
    reactions (lower set of symbols).
  {\it Right:} Energy
    dependence of the relative strangeness production as measured by
    the $E_S$ ratio (see text) in central Pb+Pb and Au+Au collisions
    (upper set of symbols) compared to results from $p+p$
    reactions (lower set of symbols).
    The compilation of data is from Ref.~\cite{evidence}.
    The dashed-dotted line in the figure
    shows the predictions of the SMES~\cite{GaGo}.
}
\end{center}
\end{figure}

The $E_S$ ratio was the first observable used to establish the energy
dependence of the strangeness to entropy production ratio from the data.
This ratio is closely proportional (see Appendix~8.2) to the
$\langle K^+ \rangle/\langle \pi^+ \rangle$ ratio, which with time
became better measured experimentally.
The energy dependence of both
ratios is plotted in Fig.~\ref{strangeness} for
central Pb+Pb (Au+Au) collisions and $p+p$ interactions
as function of collision energy.
%

For $p+p$ interactions the ratios show a monotonic increase with
energy. In contrast, a very different behavior is observed for central
Pb+Pb (Au+Au) collisions. The steep threshold rise of the ratios
characteristic for confined matter changes at high energy into a constant value
at the level expected for deconfined matter. In the transition
region (at low SPS energies) a sharp maximum is observed caused by
the higher strangeness to entropy production ratio in confined matter than
in deconfined matter. As seen in Fig.~\ref{strangeness} the
measured dependence is consistent with that predicted within the
SMES~\cite{GaGo}.

\subsection{The Step}
The energy density at the early stage increases with increasing
collision energy. At low and high energies, when a pure confined
or deconfined phase is produced, this leads to an increase of the
initial temperature and pressure. This, in turn, results in an
increase of the transverse expansion of the produced matter and consequently
a flattening of the transverse mass spectra of final state
hadrons. One may expect an `anomaly' \cite{van-hove,Hu:95,GoGaBu}
in the energy dependence of the average hadron transverse momenta
in the mixed phase region where the
temperature and pressure are approximately constant.

The experimental data on spectra of the transverse
mass $m_T= (m^2+p_T^2)^{1/2}$  are usually
parameterized by a simple exponential dependence:
\begin{equation}\label{T*}
\frac{dN}{m_Tdm_T}~\cong~C~\exp\left(-~\frac{m_T}{T^*}\right)~.
\end{equation}
The inverse slope parameter
$T^{*}$ is sensitive to both the thermal and collective motion in
the transverse direction. Hydrodynamical transverse flow with
collective velocity  $v_T$ modifies the Boltzmann $m_T$-spectrum
of hadrons. At low transverse momenta, it leads to the result
($T_{kin}$ is the kinetic freeze-out temperature):
\begin{equation}\label{T*1}
T^*_{low-p_T} ~\cong ~T_{kin}~+~\frac{1}{2} m~v_T^2~.
\end{equation}
Such a linear mass dependence of $T^*$ is supported by the data
for hadron spectra at small $p_T$. However, for $p_T~\gg~m$ the
hydrodynamical transverse flow leads to the mass-independent
blue-shifted `temperature':
\begin{equation}\label{T*2}
T^*_{high-p_T} ~=~T_{kin}~\cdot~ \sqrt{\frac{1+v_T}{1-v_T}}~.
\end{equation}
%
Note that a simple exponential fit Eq.~(\ref{T*}) neither  works for
light $\pi$-mesons, $T^{*}_{low-p_{T}}(\pi) <
T^{*}_{high-p_{T}}(\pi)$, nor for heavy (anti-)protons and
(anti-)lambdas, $T^{*}_{low-p_{T}}(p,\Lambda)
> T^{*}_{high-p_{T}}(p,\Lambda)$
(see e.g., Refs.~\cite{Teaney:2000cw,Gorenstein:2001ti}).

Kaons are the best suited among measured hadron
species for observing the effect of the modification of the
EoS due to the onset of deconfinement
in hadron transverse momentum spectra.
The arguments are the following.
First, the kaon $m_{T}$--spectra are only weakly affected by
hadron re-scattering and resonance decays during the
post-hydrodynamic hadron cascade at SPS and RHIC energies
\cite{Teaney:2000cw}.
Second,  a simple one parameter exponential fit Eq.~(\ref{T*}) is quite
accurate for kaons in central A+A collisions at all energies. This
simplifies the analysis of the experimental data.
Third, high quality data on $m_T$-spectra of $K^+$ and $K^-$
mesons in central Pb+Pb (Au+Au) collisions are available over the
full range of relevant energies.

\begin{figure}[!htb]
\begin{center}
\begin{minipage}[b]{1.0\linewidth}
\includegraphics[width=0.5\linewidth]{kap_slope_sqrts.pdf}
\includegraphics[width=0.5\linewidth]{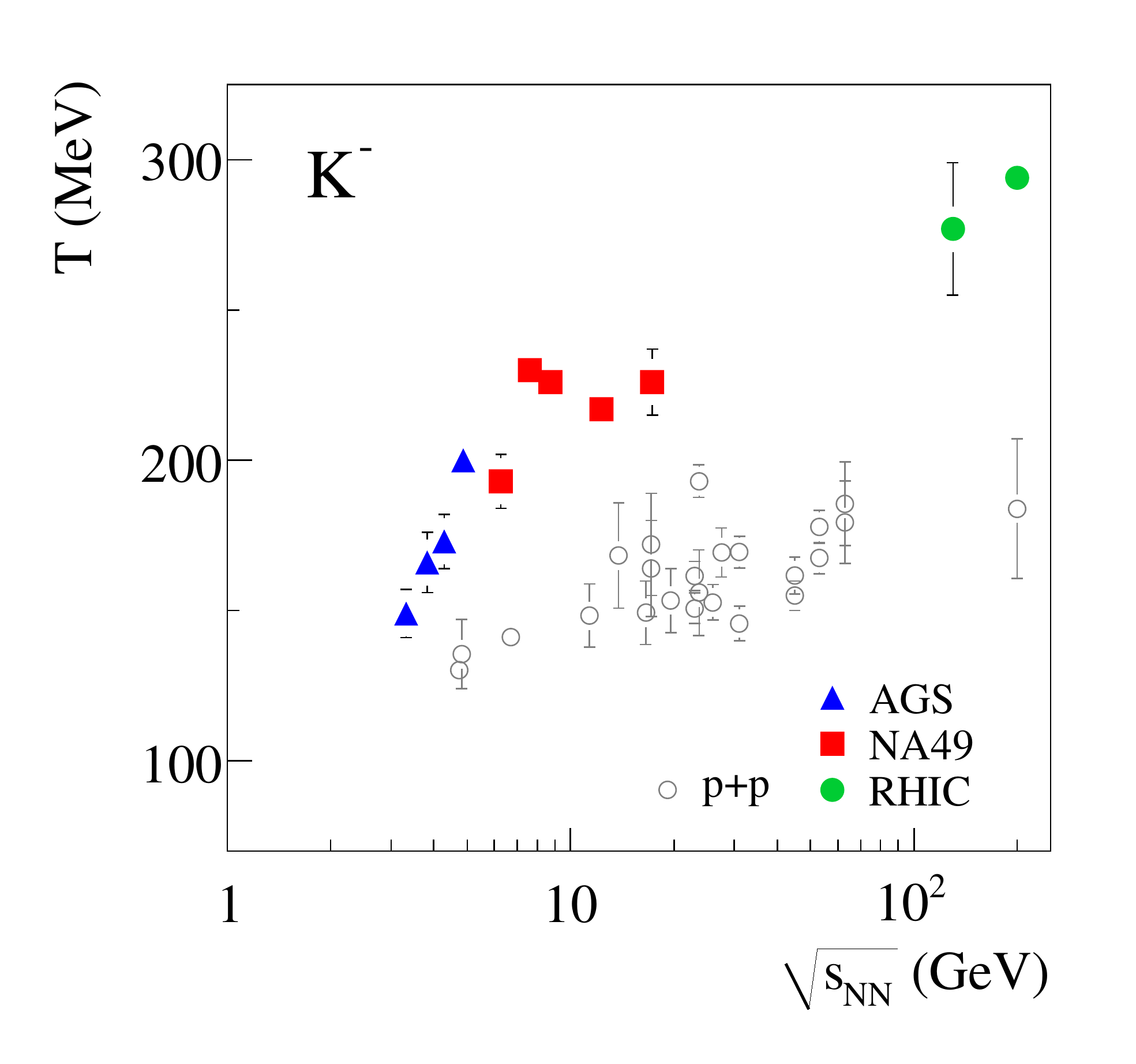}
\end{minipage}
\caption{\label{slopes}
  Energy dependence of the inverse slope parameter $T^*$ of the
  transverse mass spectra of $K^+$ ({\it left}) and $K^-$ mesons
  ({\it right})
  measured at mid-rapidity in central Pb+Pb and Au+Au
  collisions. The $K^{\pm}$ slope parameters are compared to those from
  $p +p$ reactions (open circles). The compilation of data is from
  Ref.~\cite{evidence}.
}
\end{center}
\end{figure}


\begin{figure}[!htb]
\begin{center}
\begin{minipage}[b]{1.0\linewidth}
\includegraphics[width=1.0\linewidth]{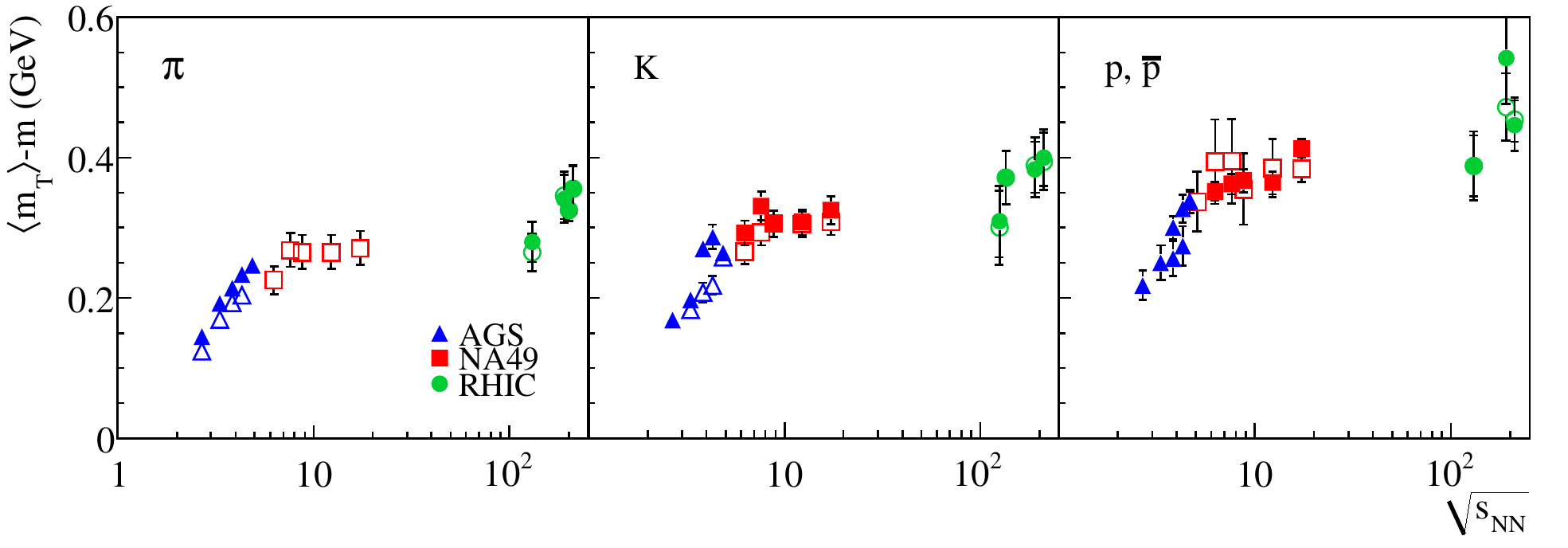}
\end{minipage}
\caption{\label{mt}
Energy dependence of the mean transverse mass,
$\langle m_T\rangle ~-~m$, measured at
mid-rapidity in central Pb+Pb and Au+Au collisions for $\pi^{\pm}$
({\it left}),  $K^{\pm}$ ({\it middle}),
and $p$ and $\bar{p}$ ({\it right}).
Results for positively (negatively) charged hadrons are
shown by full (open) symbols.  The compilation of data is
from Ref.~\cite{evidence}.
}
\end{center}
\end{figure}

The experimental results on the energy dependence of the inverse
slope parameter of $K^+$  and $K^-$ transverse mass spectra for
central Pb+Pb (Au+Au) collisions are shown in Fig.~\ref{slopes}.
The striking features of the data can be summarized and
interpreted~\cite{GoGaBu} as follows. The $T^{*}$ parameter
increases strongly with collision energy up to the SPS energy
point at $\sqrt{s}_{NN}~=$~7.6~GeV (30$A$~GeV). This is the energy
region where the creation of confined matter at the early stage of
the collisions is expected. Increasing collision energy leads to
an increase of the early stage temperature and pressure.
Consequently the transverse momenta of produced hadrons, measured
by the inverse slope parameter, increase with collision energy.
This rise is followed by a region of approximately constant value
of the $T^{*}$ parameter in the SPS energy range
$\sqrt{s}_{NN}=$~7.6~-~17.2~GeV (30$A$~-~158$A$~GeV), where one
expects the transition between confined and deconfined matter with
the creation of mixed phases. The resulting modification of the
equation of state `suppresses' the hydrodynamical transverse
expansion and leads to the observed plateau structure in the
energy dependence of the $T^*$ parameter \cite{GoGaBu}. At higher
energies (RHIC data), $T^{*}$ again increases with the collision
energy. The EoS at the early stage becomes again stiff and the
early stage pressure increases with collision energy, resulting in
a resumed increase of $T^{*}$.
%
%
%
%
%
As also shown in Fig.\ref{slopes} the parameter $T^*$ in $p+p$ interactions
appears to increase smoothly and does not show the step-like structure.

For the transverse mass spectra of pions and protons the inverse
slope parameter depends on the transverse mass interval used in
the fit.  The mean transverse mass $\langle m_T\rangle$ provides
an alternative characterization of the $m_T$-spectra.
The energy dependence of $\langle m_T\rangle~-~m$ for pions, kaons
and (anti-)protons is shown in Fig.\ref{mt}.
%
%
%
These results demonstrate that the approximate
energy independence of $\langle m_T\rangle~-~m$ in the SPS energy
range is a common feature for all abundantly produced particle species.

\subsection{The Dale}
This subsection summarizes the analysis of Ref.~\cite{bleicher} of
the longitudinal pion spectra
within Landau's hydrodynamical model \cite{Landau:53,Landau:55}.
The interest in this model was revived by the remarkable
observation that the rapidity distributions at all investigated
energies can be well described by a single Gaussian
(see \cite{Roland:2004} and references therein).
Moreover, the energy dependence of the
width can also be described reasonably well by the same model.

\begin{figure}[!htb]
\begin{center}
\begin{minipage}[b]{1.0\linewidth}
\includegraphics[width=0.5\linewidth]{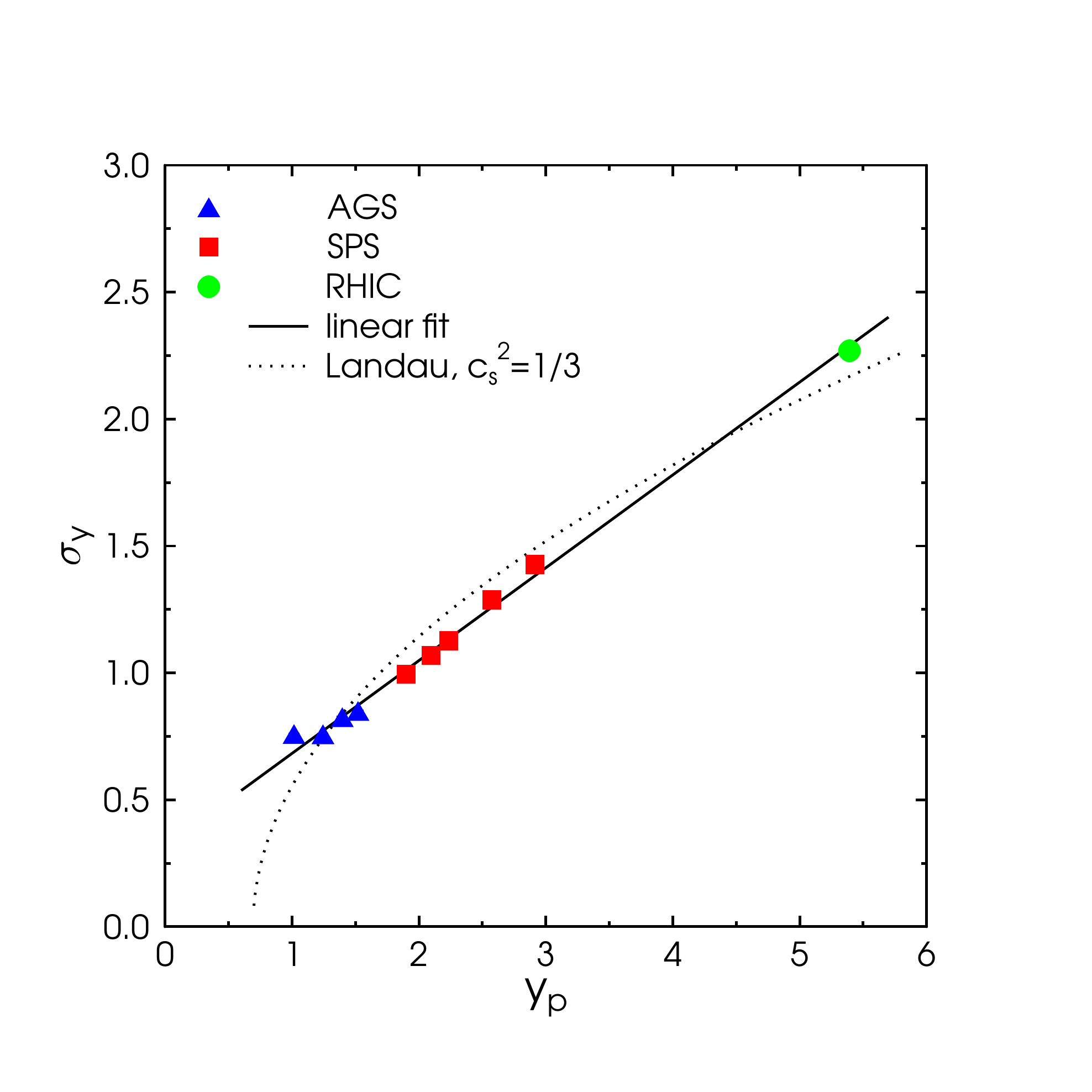}
\includegraphics[width=0.5\linewidth]{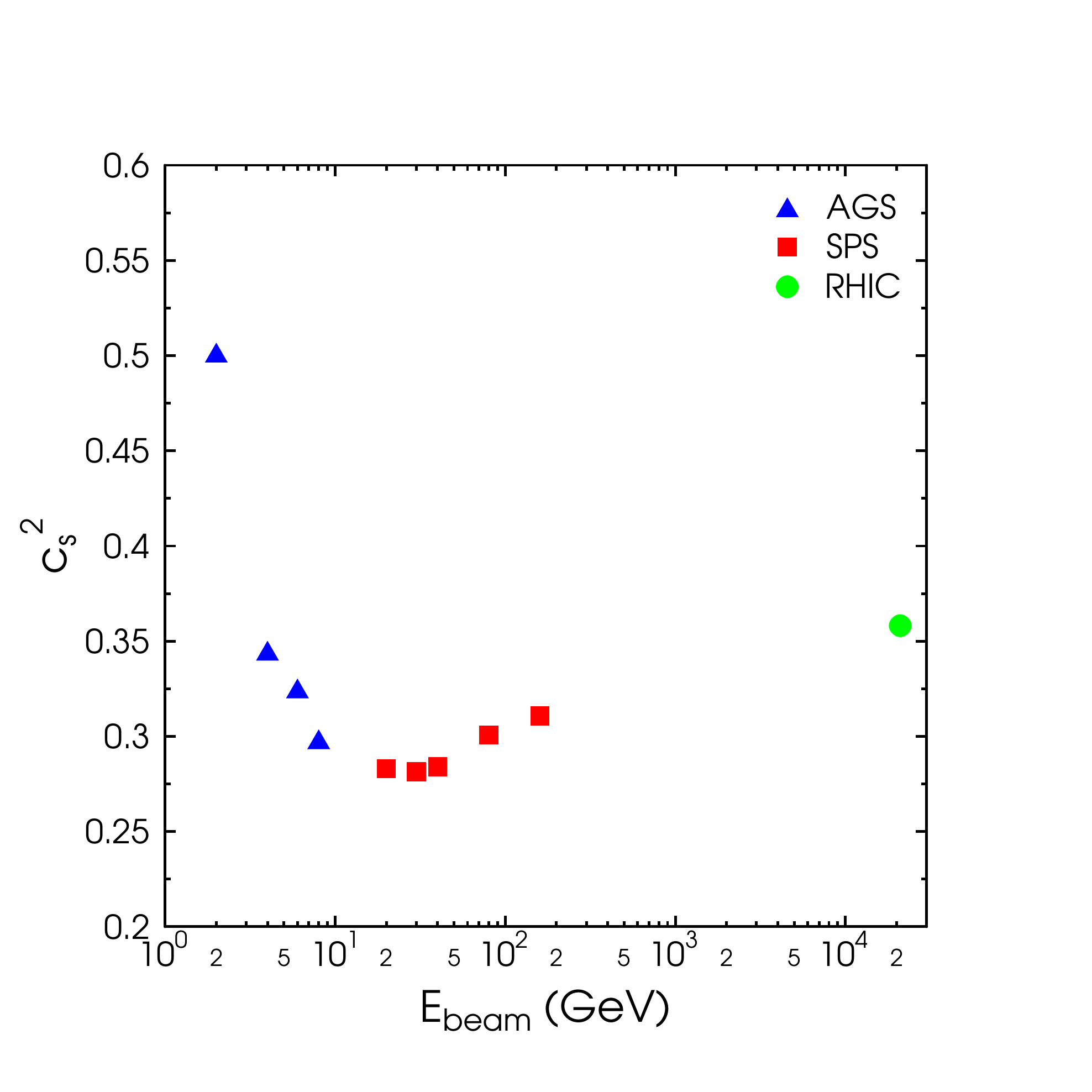}
\end{minipage}
\caption{\label{rapwidth}
Comparison of the Landau hydrodynamic model with rapidity
distributions of charged particles~\cite{bleicher}. {\it Left:} The root mean
square width $\sigma_y$ of the rapidity distributions of negatively
charged pions in central Pb+Pb (Au+Au) reactions as a function of
the beam rapidity $y_p$. The dotted line indicates the Landau
model prediction with  $c_s^2=1/3$, while the full line shows a
linear fit through the data points. Data (full symbols) are taken
from \cite{na49_blume,Roland:2004,klay,brahms}. The statistical
errors given by the experiments are smaller than the symbol sizes.
Systematic errors are not available. {\it Right:} Speed of sound
as a function of beam energy for central Pb+Pb (Au+Au) reactions
as extracted from the data using~Eq.~(\ref{eq3}). The statistical
errors (not shown) are smaller than~3\%.
}
\end{center}
\end{figure}

The main physics assumptions of Landau's picture are as follows.
The collision of two Lorentz-contracted nuclei leads to a complete
stopping of the colliding nuclei and full thermalization of the
created hadronic matter. This establishes the volume and energy
density for the initial conditions of hydrodynamic expansion at
each collision energy. Assuming for simplicity the equation of
state in the form $p=c_s^2\varepsilon$
($c_s$ denotes the speed of sound, and $c_s^2=1/3$ for an ideal
massless particle gas)  the pion rapidity spectrum is given
by~\cite{Shuryak:1972zq,Carruthers:dw}:
\begin{equation}
\frac{dn}{dy}=\frac{Ks_{\rm NN}^{1/4}}{\sqrt{2\pi
\sigma_y^2}}\,\exp\left(-\frac{y^2}{2\sigma_y^2}\right)
\label{eq1}
\end{equation}
with
\begin{equation}
\sigma_y^2=\frac{8}{3}\frac{c_s^2}{1-c_s^4}\,{\rm ln}(\sqrt
{s}_{\rm NN}/{2m_{\rm N}})\quad, \label{eq2}
\end{equation}
where $K$ is a normalization factor converting 
entropy to pion density~\footnote{
There are two issues related to derivation of
Eq.~(\ref{eq2}) which need clarification by future study.
First,  Eq.~(\ref{eq2}) is obtained assuming that $c_s$ depends
only on the early stage energy density and its dependence on
decreasing energy density during expansion is neglected.
Second, the Landau model assumes stopping and thermalization
of the total energy in the c.m. system, whereas only a fraction
of the inelastic energy is stopped and thermalized in the SMES model.}.
The above prediction was compared with the experimental data on
rapidity distributions of negatively charged pions  produced in
central Pb+Pb (Au+Au) collisions at various energies.
Figure~\ref{rapwidth}~{\it left} shows the measured width
$\sigma_y$ of the rapidity
spectra~\cite{na49_blume,Roland:2004,klay,brahms} as a function of
the beam rapidity. The full line shows a linear fit through the
data points. The dotted line indicates the Landau model
predictions with $c_s^2=1/3$.
%
%
The model only roughly reproduces the measured dependence. At low
AGS energies and at the top RHIC energy, the experimental points
are under-predicted, while in the SPS energy regime Landau's model
over-predicts the width of the rapidity distributions.

These deviations can be attributed to the changes in the EoS,
which can be effectively parameterized by allowing the speed of
sound to be dependent on collision energy. By inverting
Eq.~(\ref{eq2}) one can express $c_s^2$ in the medium as a
function of the measured width of the rapidity distribution:
\begin{equation}
c_s^2~=~-~\frac{4}{3}\frac{{\rm ln}(\sqrt {s}_{\rm NN}/{2
m_p})}{\sigma_y^2} ~+~\sqrt{\left[\frac{4}{3}\frac{{\rm ln}(\sqrt
{s}_{\rm NN}/{2 m_{\rm N}})}{\sigma_y^2}\right]^2+1}\quad.
\label{eq3}
\end{equation}

The energy dependence of the sound velocities extracted from the
data  using Eq.~(\ref{eq3}) is presented in  Fig.~\ref{rapwidth}
{\it right}. The sound velocities exhibit a clear  minimum
(usually called the softest point) around a beam energy of
$\sqrt{s}_{NN}~=~$7.6~GeV (30$A$~GeV).

As discussed previously the weakening of the transverse and
longitudinal expansion is expected within the SMES at low SPS
energies due to the onset of deconfinement which softens the EoS
at the early stage. Generally, a softening of the equation of
state was predicted as a signal for the mixed phase at the
transition energy from hadronic to partonic
matter~\cite{Hung:1994eq,Rischke:1995pe,Brachmann:1999mp}.
Therefore, we conclude that the data on  rapidity spectra of
negatively charged pions are indeed compatible with the assumption
of the onset of deconfinement at the low SPS energies.

\subsection{The Shark Fin and the Tooth}

As discussed in Sec.~3.4 the event-by-event fluctuations
of the energy used for particle production should lead to
fluctuations which are sensitive to the onset of deconfinement.

The NA49 Collaboration looked for the {\it shark fin} structure
in the energy dependence of the scaled variance of
multiplicity distributions in central
Pb+Pb collisions~\cite{na49-fin}.
The predicted~\cite{GaGoMo} increase of the scaled variance of multiplicity
distribution in the NA49
acceptance by about 0.01 due to the onset of deconfinement is
smaller than the systematic error on the measurement.
Therefore these data can neither support nor disprove
the {\it shark fin} prediction.

The {\it tooth} structure in the energy dependence of $R_{s/e}$
shown in Fig.~\ref{rf}~{\it right} might be seen in the
event-by-event fluctuations of the $K$/$\pi$ ratio. The energy
dependence of the fluctuations of this ratio in central Pb+Pb collisions
was studied by NA49 using the so-called $\sigma_{dyn}$
measure~\cite{na49-tooth}. The `dynamical' $K/\pi$ fluctuations
increase significantly with decreasing energy below 40$A$~GeV. It
is  unclear whether this increase is related to the rapid increase
of the $R_{s/e}$ measure predicted due to the onset of
deconfinement at energies below 30$A$~GeV.

\section{Alternative Approaches }

Several other analyzes of the energy dependence of hadron production
properties in central Pb+Pb and Au+Au collisions within various
theoretical approaches support the hypothesis that the onset of
deconfinement is located at low SPS energies.  In particular such
a result was obtained from studies of hadron yields within a
non-equilibrium hadron gas model~\cite{rafelski} and from an analysis
of the time evolution of the relative strangeness yield
using the momentum integrated Boltzmann equation~\cite{nayak}.
Moreover, a simultaneous analysis of the two-pion
correlation function and the transverse mass spectra found a plateau
in the averaged phase-space density at SPS energies which may be
associated with the onset of deconfinement~\cite{sinyukov}.

Recently a parity violating signal was observed in three-particle
azimuthal correlations at RHIC~\cite{star_parity}. Such an effect
was predicted~\cite{parity} since metastable domains may form in a
QGP where parity and time-reversal symmetries are locally
violated. The effect is expected to disappear when no QGP is
produced in the collisions. It can therefore serve as an another
indicator for the onset of deconfinement.

Numerous models have been developed to explain hadron production
in reactions of heavy nuclei without explicitly invoking a
transient QGP phase.  The simplest one is the statistical hadron
gas model~\cite{Ha:94} which assumes that the hadrochemical
freeze-out creates a hadron gas in equilibrium.  The temperature,
the baryon chemical potential, and the hadronization volume are
free parameters of the model and are fitted to the data at each
energy.  In this formulation, the hadron gas model cannot make any
prediction about the energy dependence of hadron production so
that an extension of the model was proposed, in which the values
of the temperature and baryon chemical potential evolve smoothly
with collision energy~\cite{Cl:01}. By construction (fit to the
energy dependence), the prevailing trend in the data is reproduced
by the model but important details are not, e.g. the decrease of
the $\langle K^+ \rangle /\langle \pi^+\rangle$ ratio between
$\sqrt{s}_{NN}~=$~7.6~and~12.3~GeV (30$A$ and 80$A$~GeV) is not
well described. The measured ratio of strangeness to pion yield in
central Pb+Pb collisions at $\sqrt{s}_{NN}~=$~17.2~GeV
(158$A$~GeV) is about 25\% lower than the expectation for the
fully equilibrated hadron gas~\cite{Cl:01,Be:98}. Two strategies
are followed in order to improve the quality of the hadron gas
model fits. First, additional parameters have been introduced
which allow for deviations from
equilibrium~\cite{Be:03,Be:05,rafelski}. Obviously the
non-equilibrium hadron gas models~\cite{Be:03,Be:05,rafelski} with
all the parameters fitted separately to the data at each energy
describe the experimental results significantly better. Secondly
the equilibrium model was extended to include hypothetical high
mass resonance states~\cite{pbm}. Again by adding additional free
parameters (mass dependence of the resonance state density and
their branching ratios) the fit quality can be improved.
Interestingly, the energy dependence of the parameters obtained
within the extended hadron gas models is
interpreted~\cite{rafelski,pbm}
 as an indication for
the onset of deconfinement at $\sqrt{s}_{NN}~\approx~$~7.6~GeV (30$A$~GeV).

Dynamical models of nucleus-nucleus collisions, such as
RQMD~\cite{RQMD}, UrQMD~\cite{URQMD} and HSD~\cite{HSD} treat the
initial nucleon-nucleon interactions within a string-hadronic
framework. In addition these models include effects such as
string-string interactions and hadronic re-scattering which are
expected to be relevant in nucleus-nucleus collisions.
RQMD~\cite{RQMD,RQMD1}, UrQMD~
\cite{URQMD,URQMD1,Bratkovskaya:2004kv} and
HSD~\cite{Bratkovskaya:2004kv} models, like the hadron gas model,
fail to describe the rapid change of hadron production properties
with collision energy in the low SPS energy range.

It was shown that the maximum in relative strangeness production
can be reproduced by invoking an unusually long lifetime of the
fireball at low SPS energies which decreases with collision
energy~\cite{tomasik}. This assumption is however difficult to
justify by dynamical models of the collision
process~\cite{URQMD,HSD}, and conflicts with the measured energy
dependence of the two-pion correlation function~\cite{hbt,hbt1}.

The step-like structure in the energy dependence of the inverse slope
parameter of the transverse mass spectra was obtained within the
hydrodynamical model by introduction of a rapid change of the
freeze-out conditions at low SPS energies~\cite{ivanov}.  However,
this assumption does not explain the increase of the inverse slope parameter
suggested by the RHIC results.

In summary, one is led to conclude that models
which do not invoke the onset of deconfinement at low SPS energies
cannot explain comprehensively and consistently the energy dependence
of hadron production properties in central Pb+Pb (Au+Au) collisions.

\section{Open problems}

Open questions related to the onset of deconfinement are
discussed in this section. First the theoretical questions
are addressed, then the experimental issues are discussed.

\subsection{Theoretical problems}

Quantum chromodynamics, the commonly accepted theory of
strong interactions,  was developed to model the interactions of
quarks and gluons as well as their color neutral composites, the
hadrons. Thus, in principle, this theory should be able to predict
whether and via which observables the onset of deconfinement can
be experimentally observed in nucleus-nucleus collisions. In QCD
the strength of the strong force between two quarks increases with
their distance. This property of the theory has as a consequence
that, in general, predictions are either very difficult or
impossible to calculate. Presently, there are no quantitative QCD
predictions yet concerning the onset of deconfinement in
nucleus-nucleus collisions.

The SMES model~\cite{GaGo}
assumes statistical particle production at the early stage of
nucleus-nucleus collisions.
%
The data on nucleus-nucleus collisions at RHIC, in particular
results on anisotropic flow, seem to require a large degree of
equilibration at the early stage of collisions~\cite{heinz}. Thus
the assumption of statistical particle production received
independent experimental support.

The additional assumptions of the SMES, which lead to
the {\it kink} and {\it horn} predictions, concern entropy and
strangeness conservation during expansion and freeze-out.
In fact, the model predictions remain at least qualitatively
unchanged if one assumes an approximate
proportionality of the final state
entropy and strangeness to their early stage values.
There is no easy way to prove or disprove these weaker
requirements.

The predictions concerning the {\it step} and {\it dale}
require assumptions concerning the equation of state of strongly
interacting matter. In particular, they rely on a general
feature of the EoS, i.e. the existence of the softest point when
the transition between QGP and confined matter occurs.
Thus, the qualitative predictions are independent of the
nature of the transition (e.g. cross-over or
1$^{st}$ order phase transition, full or partial energy
stopping and thermalization).
However, the quantitative predictions are sensitive
and a consistent description of the hydrodynamical
evolution has not been achieved yet. In particular,
the bag model equation of state with a 1$^{st}$ order phase transition
to hadron gas leads to a significant over-prediction of
transverse flow~\cite{hama}. Further studies are needed.

The predictions concerning the {\it tooth} and {\it shark fin}
are derived assuming that the early stage
volume fluctuations can be neglected for central collisions.
It is unclear, to which extent this condition is consistent
with recent attempts to describe multiplicity distributions
and high transverse momentum spectra in p+p interactions
by a statistical model with volume fluctuations~\cite{volume}.
Further studies are needed.

\subsection{Experimental issues}

The experimental results indicating the onset of deconfinement
were obtained mainly by the NA49 experiment at the CERN SPS.
Clearly, a confirmation of these measurements is necessary.

A beam energy scan program at the BNL RHIC has recently started.
Pilot results at $\sqrt{s}_{NN}=$~9.2~and~20~GeV are in agreement
with the corresponding NA49 data~\cite{rhic_sps}.
New RHIC data being taken by the STAR experiment in 2010
will allow a more conclusive verification of the NA49 results.

At the CERN SPS the new experiment NA61 started a
two dimensional system size and beam energy scan in 2009,
which will continue over the next 4 years. The measurements aim to
verify the existence of the onset of deconfinement in collisions of
medium size nuclei. Moreover, they will allow to study the
expected disappearance of the signals in collisions of light nuclei.

\section{Summary and Conclusions}

In this review we present the experimental and theoretical status
of the evidence for the threshold of quark-gluon plasma creation
in high energy nucleus-nucleus interactions. The location in
energy of this so-called onset of deconfinement, as well as key
experimental signals were predicted by the statistical model of
the early stage of the collision process~\cite{GaGo}. These
signals were searched for and observed within the energy scan
program of the NA49 Collaboration at the CERN SPS. Together with
measurements at lower (LBL, JINR, SIS, BNL AGS) and higher (BNL
RHIC) energies the properties of hadron production in heavy ion
collisions were established in a broad energy range. Their energy
dependence led to the conclusion that the  predicted signals of
the onset of deconfinement appear in a common energy domain
covered by the SPS at CERN. These features of the data serve as
strong experimental evidence for the existence of the onset of
deconfinement and thus for the quark-gluon plasma itself.

Quantitative model predictions, discussed in this review, are
derived within the statistical approach to particle production in
high energy collisions. The use of this approach has a two-fold
justification. First, it naturally includes the concept of phases
of strongly interacting matter and the transition between them.
Second, it is successful in describing numerous features of the
experimental data. The relation between the statistical approach
and the commonly accepted theory of strong interactions,  QCD,
remains unclear. This is because QCD
is difficult or impossible to evaluate  in the
energy region relevant for multi-particle production in general
and for the phase transitions of strongly interacting
matter in particular.

New experimental programs have started at the CERN SPS and
BNL RHIC which are devoted to the study of nucleus-nucleus collisions
in the energy region where the NA49 experiment found evidence
for the onset of deconfinement.
The STAR experiment at RHIC will provide a necessary confirmation
of these results.
The new CERN experiment NA61 will address the questions
how this observed phenomenon depends on the volume of matter
and what the properties of the transition region are.

\newpage
\section{Appendices}

\subsection{ Strangeness Enhancement and $J/\psi$ Suppression}

The idea of strangeness enhancement as a quark-gluon plasma
signal in nucleus-nucleus (A+A) collisions was formulated a long
time ago \cite{Rafelski}. It was based on the estimate that the
strangeness equilibration time in the QGP is of the same order
($\approx 10$ fm/c) as the expected life time of the fireball
formed in A+A collisions. Thus in the case of QGP creation
strangeness is expected to approach its QGP equilibrium value.
This equilibrium value is significantly higher than the level of
strangeness production in nucleon--nucleon (N+N) collisions.
Strangeness production in secondary hadronic interactions was
estimated to be negligibly small.
Therefore, if QGP is not formed, strangeness yields
would be expected to be much lower than those predicted by
equilibrium QGP calculations. Thus at that time a simple and
elegant signature of QGP creation appeared: a transition to QGP
should be signaled by an increase of the strangeness production
level to the QGP equilibrium value.

\begin{figure}[!htb]
\begin{center}
\begin{minipage}[b]{0.7\linewidth}
\includegraphics[width=1.0\linewidth]{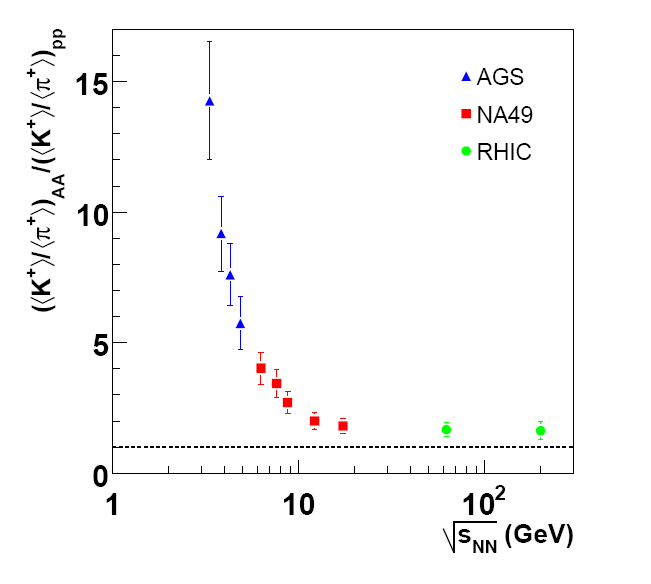}
\end{minipage}
\caption{\label{enhancement} Energy dependence of strangeness
enhancement in central Pb+Pb (Au+Au) collisions~\cite{mitrovski}.
}
\end{center}
\end{figure}

The  actual study
has been done in the following way.
The strangeness to pion ratio quantified by the ratios,
\begin{equation}\label{strange}
E_S~=~\frac{\langle \Lambda\rangle ~+~\langle
K+\overline{K}\rangle} {\langle \pi \rangle}~~~~ {\rm or}~~~~
\frac {\langle K^+\rangle } {\langle \pi^+\rangle }~ ,
\end{equation}
was measured and analyzed. One expected  that the ratios should
increase {\it strongly} in A+A collisions if the QGP was formed.
To reveal  the specific increase of the strangeness/pion ratio in
A+A collisions due to QGP formation the strangeness enhancement
factor was introduced:
\begin{equation}\label{senh}
R_S(\sqrt{s}_{NN})
~\equiv~\frac{E_S^{AA}(\sqrt{s}_{NN})}{E_S^{NN}(\sqrt{s}_{NN})}
\approx \frac{\left(\langle K^+\rangle /\langle \pi^+\rangle
\right)_{AA}}{\left(\langle K^+\rangle /\langle \pi^+\rangle
\right)_{NN}} ~,
\end{equation}
where superscripts $^{AA}$ and $^{NN}$ correspond respectively to
A+A and N+N collisions at the same N+N c.m.energy $\sqrt{s}_{NN}$.
The confrontation of this expectation with data was for the first
time possible in 1988 when results from $^{32}$S and $^{28}$Si
beams at the SPS and the AGS became available. Experiment NA35
reported that in central S+S collisions at 200$A$~GeV the
strangeness to pion ratio is indeed about 2 times higher than in
N+N interactions at the same energy per nucleon~\cite{na35}. But
an even larger enhancement ($R_S$ is about 5)  was measured by
E802 in Si+A collisions at the AGS. The data on central Au+Au
collisions at low AGS energies 2$A$--10$A$~GeV completed the
picture: strangeness enhancement is observed at all energies, and
it is stronger  at lower energies, i.e. the ratio $R_s$
Eq.(\ref{senh}) {\it increases} monotonically with {\it
decreasing}  $\sqrt{s}_{NN}$. Figure~\ref{enhancement} shows a
compilation of recent data~\cite{mitrovski}.

Moreover, the enhancement factor $R_s$ Eq.(\ref{senh}) should
evidently go to infinity at the threshold energy of strange hadron
production in N+N collisions. At low AGS energies one does not
expect the creation of a QGP and therefore the substantial
strangeness enhancement is clearly of a different origin. The low
level of strangeness production in p+p interactions as compared to
the strangeness yield in central A+A collisions can be understood
to a large extent in the statistical model as due to the effect of
exact strangeness conservation. The canonical ensemble treatment
of strangeness conservation leads to additional suppression
factors imposed on strange hadron production in small systems such
as created in p+p collisions. In any case, the AGS measurements
indicating a strangeness enhancement $R_s$ larger than that at the
SPS show clearly that the simple concept of strangeness
enhancement as a signal of QGP production does not work.

\vspace{0.3cm} The standard picture of $J/\psi$ production in
collisions of hadrons and nuclei assumes a two step process: the
creation of a $c\overline{c}$ pair in hard parton collisions at
the very early stage of the reaction and subsequent formation of a
bound charmonium state. It was proposed~\cite{Satz} to use the
$J/\psi$ as a probe for deconfinement in the study of A+A
collisions. The argument was that in a QGP color screening
dissolves initially created $J/\psi$ mesons into $c$ and
$\overline{c}$ quarks which at hadronization form open charm
hadrons. As the initial yield of $J/\psi$ is believed to have the
same A--dependence as the yield of Dell--Yan lepton pairs, the
measurement of a weaker A--dependence of the final $J/\psi$ yield
($J/\psi$ suppression) would signal charmonium absorption and
therefore creation of QGP.

Production of charmonium states $J/\psi$ and $\psi^{\prime}$ was
measured in A+A collisions at the CERN SPS over a period of 15
years by the NA38, NA50 and NA60 Collaborations~\cite{jpsi}. The
A-dependence of $J/\psi$ production in p+A is weaker than A$^1$
(approximately A$^{0.9}$). It was suggested that this $J/\psi$
suppression is due to absorption on nucleons in the target
nucleus. The data on oxygen and sulfur collisions on nuclei at
200$A$~GeV also indicated considerable suppression. To improve the
fit of the data a new source of $J/\psi$ absorption was
introduced: the absorption on hadronic secondaries (`co-movers').
Finally in central Pb+Pb collisions at 158$A$~GeV the measured
suppression was found to be significantly stronger than expected
in the models including nuclear and co-mover suppression. This
`anomalous' $J/\psi$ suppression was interpreted as evidence of
QGP creation in Pb+Pb collisions at the CERN SPS. The
uncertainties in estimates of the $J/\psi$ absorption by target
nucleons and co-movers make $J/\psi$ suppression a problematic QGP
signal. An essential part of the $J/\psi$ yield comes from decays
of excited charmonium states like $\psi^{\prime}$ and $\chi$. All
of them have different melting temperatures and absorption
cross-sections.

Alternative approaches have been developed, namely the statistical~\cite{Ga1}
and the statistical coalescence~\cite{Br1,Go:00}
models of $J/\psi$ production, which
reproduce the A-dependence of the $J/\psi$
yield at SPS energies reasonably well.
They are based on different physics pictures than
the one leading to $J/\psi$ suppression as the signal
of quark-gluon plasma creation. Specifically,
the statistical model~\cite{Ga1} assumes statistical
production of $J/\psi$ mesons at hadronization, whereas
in the coalescence model
statistical coalescence of $c$ and $\overline{c}$ quarks
at hadronization is assumed ~\cite{Br1,Go:00}.
In both models the $J/\psi$ yield  is neither related to the J/psi
suppression in the hadron gas nor in the quark-gluon plasma.

\newpage

\subsection{ Main carriers of $s$ and $\bar{s}$ quarks}
\begin{figure}[!htb]
\begin{center}
\begin{minipage}[b]{1.0\linewidth}
\includegraphics[width=1.0\linewidth]{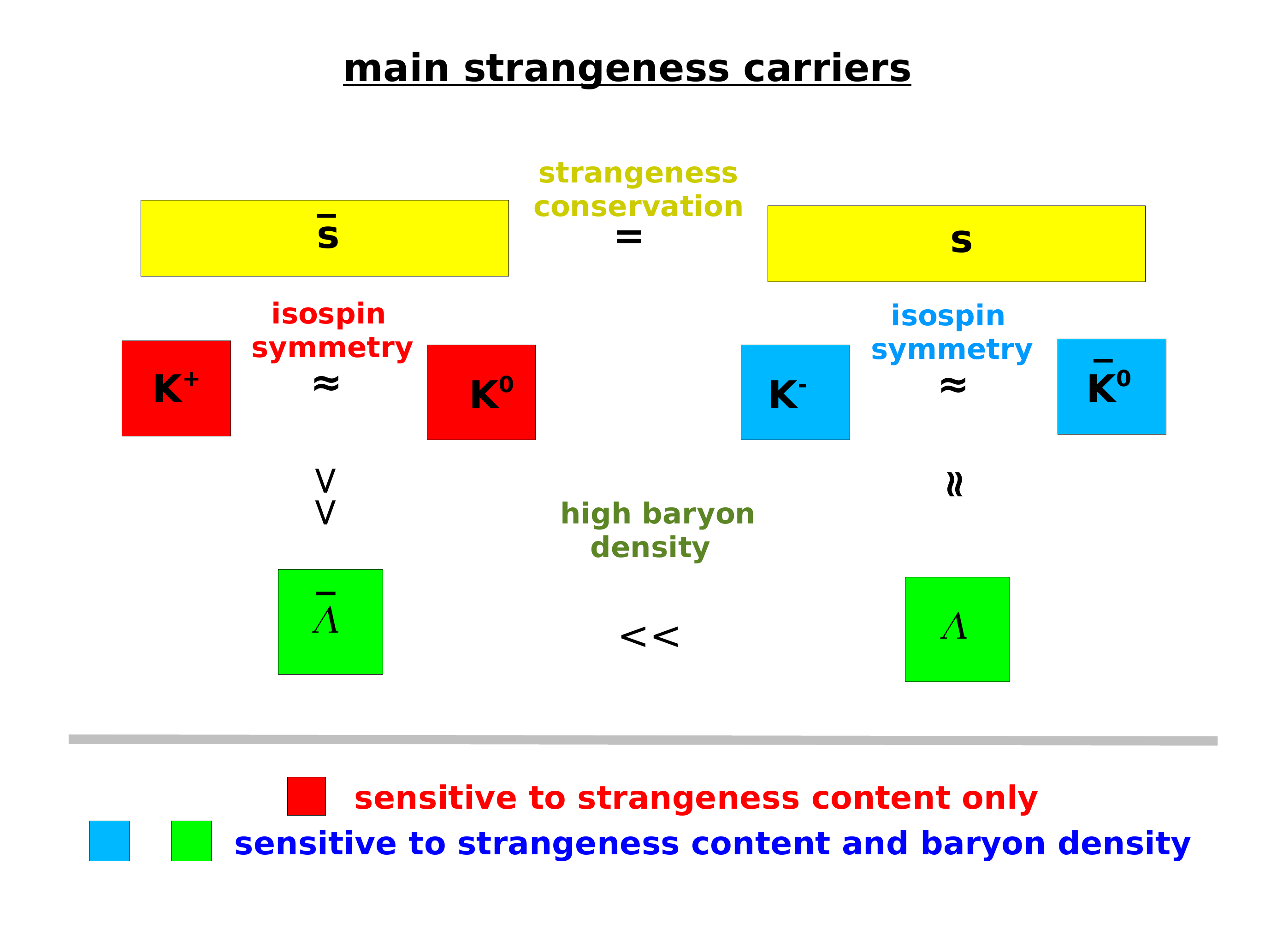}
\end{minipage}
\caption{\label{strangeness_carriers}
Main carriers of $s$ and $\bar{s}$ quarks.
$K^-$ and $\bar{K}^0$ as well as $\Lambda$ and
$\bar{\Lambda}$ yields are sensitive
to the strangeness content and baryon density.
$K^+$ and $K^0$ yields are sensitive mainly to strangeness content.
}
\end{center}
\end{figure}
The distribution of $s$ and $\bar{s}$ quarks among the most
abundantly produced hadrons is considered  here for the case of nucleus-nucleus
collisions at the SPS energies.
The sketch presented in Fig.~\ref{strangeness_carriers} illustrates
the following discussion.

The colliding nuclei have net numbers of $s$ and $\bar{s}$ quarks
equal to zero. As strangeness is conserved in strong interactions
the numbers of $s$ and $\bar{s}$ quarks in the final state
have to be equal.

Kaons are the lightest strange hadrons. The isospin partners $K^+$
and  $K^0$ mesons carry $\bar{s}$ quarks, whereas $K^-$ and
$\bar{K}^0$ carry $s$ quarks. The (approximate) symmetry of the
initial state and isospin conservation in strong interactions
imply that:
\begin{equation}
\langle K^+\rangle \approx \langle K^0\rangle
\end{equation}
and
\begin{equation}
\langle K^-\rangle \approx \langle \bar{K}^0\rangle .
\end{equation}
The $\overline{s}$-quarks are also carried by the lightest
anti-baryon, $\bar{\Lambda}$. Its fraction is however small (less
than 5\%) at the AGS and SPS energies due to suppression of the
anti-baryon yield by the high net-baryon density. Consequently,
$K^+$ and $K^0$ mesons carry each about half of all the
anti-strange quarks produced in A+A collisions at AGS and SPS
energies. Thus, their yields are nearly proportional to the total
number of produced $s$ and $\bar{s}$ quarks.

This is not the case for $K^-$ and $\bar{K}^0$ mesons.
A significant fraction of $s$-quarks (about 50\% in central Pb+Pb
collisions at 158~A$\cdot$GeV) is carried by hyperons.
In addition this fraction strongly depends on collision energy.
Consequently, the fraction of $s$-quarks carried by
anti-kaons, $K^-$ and $\overline{K}^0$, is also
dependent on collision energy and cannot be used easily
to quantify strangeness production.
In the $E_S$ ratio all main carriers of strange and anti-strange
quarks are included. The neglected contribution of
$\overline{\Lambda}$ and other hyperons and anti-hyperons is about
10\% at SPS energies.  Both the $\langle K^+\rangle /\langle
\pi^+\rangle$ and $E_S$ ratios are approximately, within 5\% at
SPS energies, proportional to the ratio of total multiplicity of
$s$ and $\overline{s}$ quarks to the multiplicity of pions.  It
should be noted that the $\langle K^+\rangle/\langle \pi^+\rangle$
ratio is expected to be similar (within about 10\%) for $p+p$,
$n+p$, and $n+n$ interactions at 158$A$~GeV, whereas the $E_S$
ratio is independent of the isospin of nucleon-nucleon
interactions.

\newpage
\subsection{ Onset of deconfinement and critical point}
\begin{figure}[!htb]
\begin{center}
\begin{minipage}[b]{1.0\linewidth}
\includegraphics[width=1.0\linewidth]{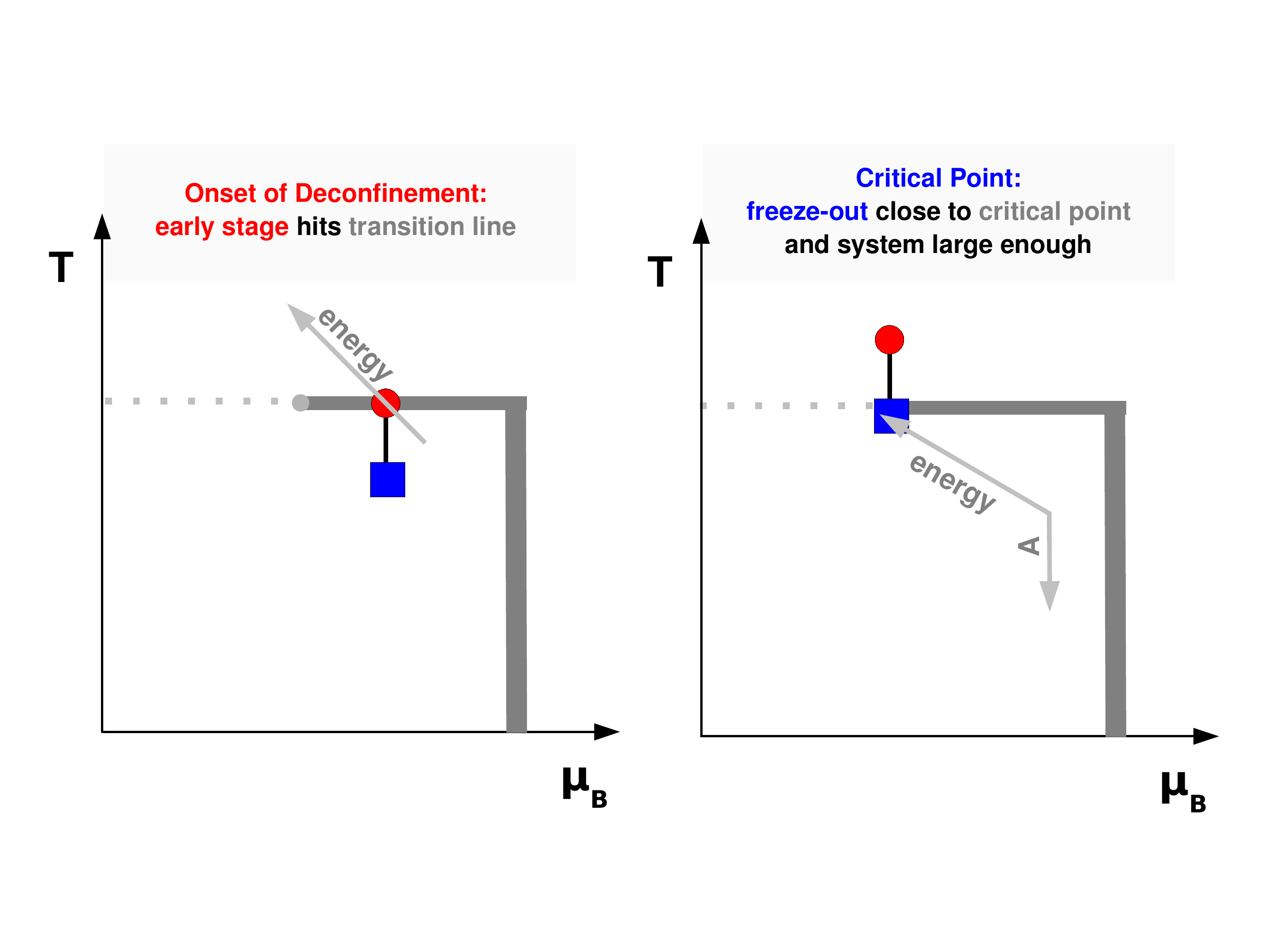}
\end{minipage}
\caption{\label{cpod}
Onset of deconfinement and critical point.
}
\end{center}
\end{figure}
This appendix discusses relations between the onset of deconfinement,
the critical point of strongly interacting matter and the
possibilities of their experimental study in relativistic nucleus-nucleus
collisions. The two sketches presented in Fig.~\ref{cpod}
should help to understand the basic ideas.

The onset of deconfinement refers to the beginning of
the creation of a deconfined state of strongly interacting matter
(ultimately a quark-gluon plasma)  at the early stage of
nucleus-nucleus collisions when increasing the collision energy.
With increasing collision energy the energy density of matter created
at the early stage of A+A collisions increases.
Thus, if there are two phases\footnote{
The discussed two phase diagram is the simplest one
which allows to introduce the concepts  of the
onset of deconfinement and the transition region.
There are numerous suggestions of phase diagrams with a
significantly richer structure (see e.g., Ref.~\cite{diagram}).}
of matter separated by the transition
region (solid and dotted lines)  as indicated in Fig.~\ref{cpod}~{\it left}
the early stage (the red point) first has to  hit and then
move above the transition region.
Therefore, the existence of the onset of deconfinement is
the most straightforward consequence of the existence of
two phases of strongly interacting matter, i.e. confined matter
and QGP.
The experimental observation of the onset of deconfinement
required a one dimensional scan in collision energy with
heavy ions as performed by NA49.
All signals of the onset of deconfinement discussed in this
paper relate to the difference in properties of confined
matter and QGP. They are not sensitive to the structure of the
transition region.

Discovery of the onset of deconfinement implies the existence of
QGP and of a transition region between confined and QGP phases.
Numerous possibilities concerning the structure of the transition
region are under discussion (see e.g., Ref.~\cite{kapusta}). The
most popular one~\cite{ssr}, sketched in Fig.~\ref{cpod}, claims
that a 1$^{st}$ order phase transition (thick gray line) separates
both phases in the high baryonic chemical potential domain. In the
low baryonic chemical potential domain a rapid crossover is
expected (dotted line). The end point of the 1$^{st}$ order phase
transition line is the critical point.

The characteristic signatures of the critical point can be
observed if the freeze-out point (blue square in
Fig.~\ref{cpod}~{\it right}) is located close to the critical
point. The analysis of the existing experimental data~\cite{Be:05}
indicates that the location of the freeze-out point in the phase
diagram depends on the collision energy and the mass of the
colliding nuclei. This dependence is schematically indicated in
Fig.~\ref{cpod}~{\it right}. Thus the experimental search for the
critical point requires a two-dimensional scan in collision energy
and size of the colliding nuclei. The NA61
experiment~\cite{Gazdzicki:2006fy,proposal} at the CERN SPS
started this scan in 2009. It should be completed within several
years. Note, that a two dimensional scan is actually required for
any study of the structure of the transition region, independent
of the hypothesis tested.

The transition region can be studied experimentally in
nucleus-nucleus collisions only at $T$,$\mu_B$ values which
correspond to collision energies higher than the energy of the
onset of deconfinement. This important conclusion is easy to
understand when looking at Fig~\ref{cpod}. Signals of the critical
point can be observed provided the freeze-out point is close to it
(see Fig.~\ref{cpod}~$right$). On the other hand, by definition
the critical point is located on the transition line. Furthermore,
the energy density at the early stage of the collision is, of course,
higher than the energy density at freeze-out. 
Thus, the condition that the freeze-out point is near the
critical point implies that the early stage of the system
is above (or on) the transition line. This in turn means that 
the optimal energy range for the search for the critical point
lies above the energy of the onset of deconfinement 
(see Fig.~\ref{cpod}~$left$).
This general condition limits the search for the critical point to
the collision energy range $E_{LAB} > 30A$~GeV.


\end{document}